\documentclass[12pt]{article}
\usepackage{graphicx,amsmath,amssymb}
\usepackage{indentfirst}
\usepackage[usenames]{color}
\usepackage[colorlinks=true, urlcolor=navyblue, linkcolor=navyblue, citecolor=navyblue]{hyperref}
\usepackage{epstopdf}
\usepackage{enumerate}
\usepackage{appendix}
\usepackage{lineno}
\usepackage{setspace}

\definecolor{navyblue}{rgb}{0,0.08,0.45}
\definecolor{darkred}{rgb}{0.7,0.0,0.0}
\definecolor{darkgreen}{rgb}{0,0.6,0.2}

\newcommand{\beq}{\begin{equation}}
\newcommand{\enq}{\end{equation}}
\newcommand{\beqa}{\begin{eqnarray}}
\newcommand{\beqast}{\begin{eqnarray*}}
\newcommand{\enqa}{\end{eqnarray}}
\newcommand{\enqast}{\end{eqnarray*}}

\newcommand{\bec}{\begin{center}}
\newcommand{\enc}{\end{center}}
\newcommand{\beqo}{\begin{quote}}
\newcommand{\enqo}{\end{quote}}
\newcommand{\bem}{\begin{minipage}}
\newcommand{\enm}{\end{minipage}}

\begin{document}

\vspace{15pt}

\begin{center}
{\large Hadron Spectroscopy and Dynamics from Light-Front Holography and Superconformal Algebra}

\vspace{10pt}


\end{center}

\vspace{15pt}

\centerline{Stanley J. Brodsky}

\vspace{3pt}

\centerline {\it SLAC National Accelerator Laboratory, Stanford University}

\vspace{10pt}

\begin{abstract}
QCD is not supersymmetrical in the traditional sense -- the QCD Lagrangian is based on quark and gluonic fields, not squarks nor gluinos. However, its hadronic eigensolutions conform to a representation of superconformal algebra, reflecting the underlying conformal symmetry of chiral QCD and its Pauli matrix representation.   
The  eigensolutions  of superconformal algebra provide a unified Regge spectroscopy of meson, baryon, and tetraquarks of the same parity and twist  as equal-mass members of the same 4-plet representation with a universal Regge slope.  The pion $q \bar q$ eigenstate has zero mass for $m_q=0.$  The superconformal relations also can be extended to 
heavy-light quark mesons and baryons.  The combined approach of light-front holography and superconformal algebra also provides insight into the origin of the QCD mass scale and color confinement.  A key observation is the remarkable dAFF principle which shows how a mass scale can appear in the Hamiltonian and the equations of motion while retaining the conformal symmetry of the action.  When one applies the dAFF procedure to chiral QCD, a mass scale $\kappa$ appears which determines universal Regge slopes, hadron masses in the absence of the Higgs coupling, and the mass parameter underlying the Gaussian functional form of the nonperturbative QCD running coupling:  
$\alpha_s(Q^2) \propto \exp{-(Q^2/4 \kappa^2)}$,  in agreement with the effective charge  determined from measurements of the Bjorken sum rule.
The mass scale $\kappa$ underlying hadron masses  can be connected to the parameter   $\Lambda_{\overline {MS}}$ in the QCD running coupling by matching its predicted nonperturbative form to the perturbative QCD regime. The result is an effective coupling $\alpha_s(Q^2)$  defined at all momenta.   One also obtains empirically viable predictions for spacelike and timelike hadronic form factors, structure functions, distribution amplitudes, and transverse momentum distributions.  
\end{abstract}

\section{Introduction}
\label{intro}

\begin{figure}
 \begin{center}
\includegraphics[height=12cm,width=16cm]{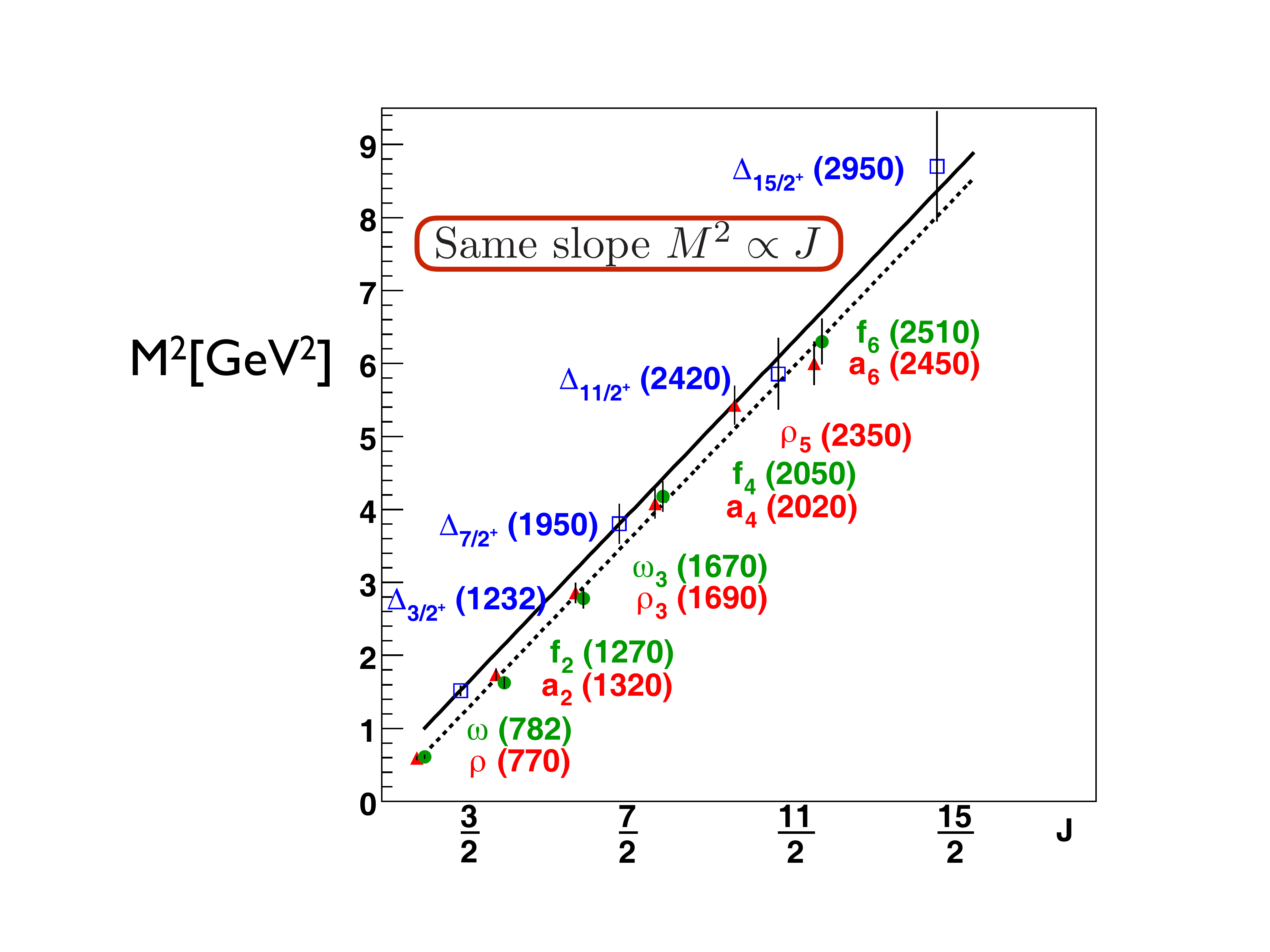}
\label{NSTARFigA.pdf}
\end{center}
\caption{Comparison of the $\rho$ mesonic and $\Delta$ baryonic Regge trajectories by Klempt and Metsch~\cite{Klempt:2012fy}. }
\end{figure}

\begin{figure}
 \begin{center}
\includegraphics[height=12cm,width=16cm]{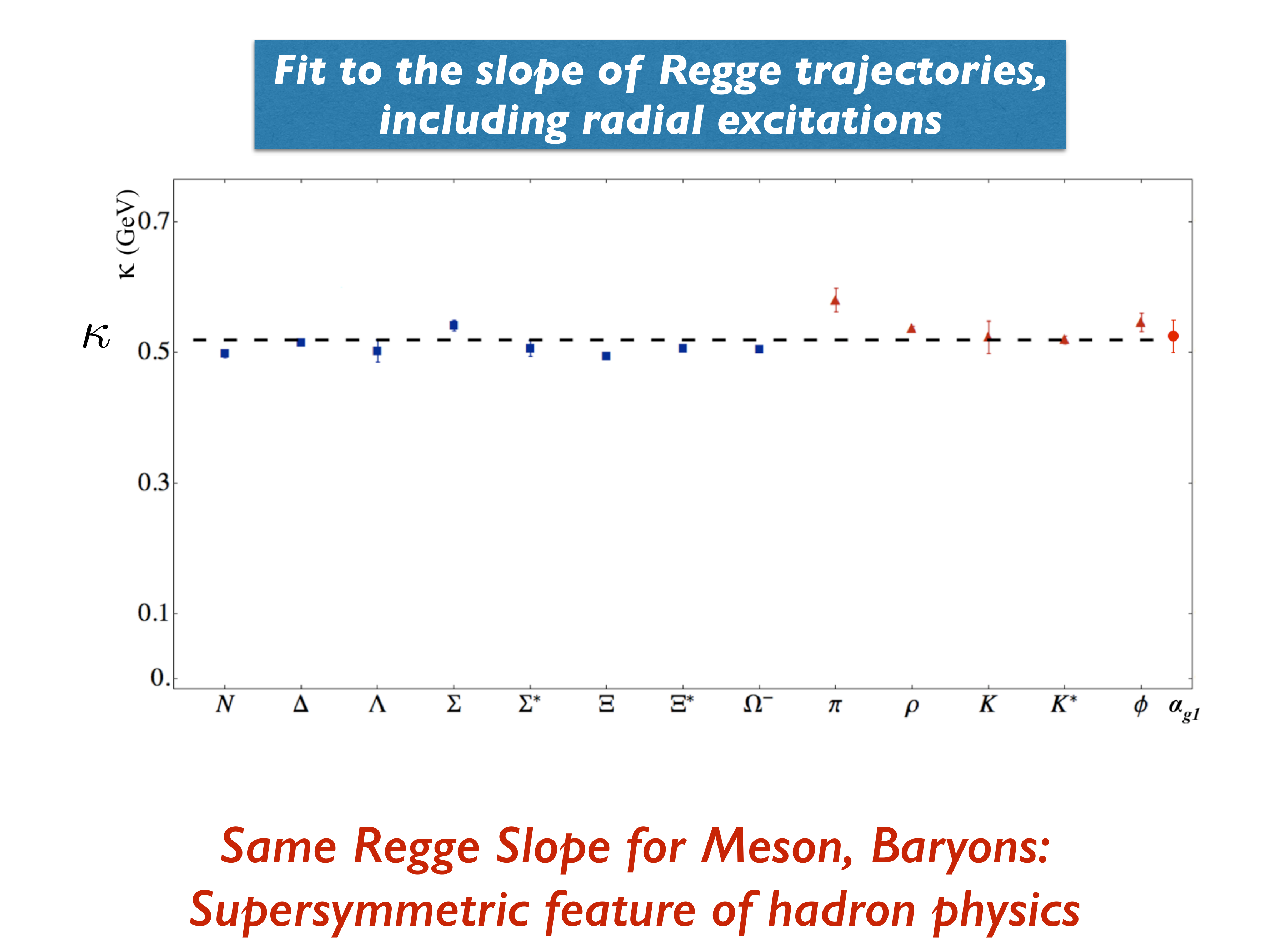}
\label{NSTARFigC}
\end{center}
\caption{Comparison of the slopes of the Regge trajectories  in angular momentum:  $M^2 \propto L$ and in the radial quantum number $n$: $M^2 \propto n$.}
\end{figure}

One of the surprising features of hadron spectroscopy is the remarkable similarity between meson and baryon Regge trajectories illustrated in Fig. \ref{NSTARFigA.pdf}.  
As shown by Klempt and Metsch~\cite{Klempt:2012fy}, the slopes for the $\rho/\omega$ and $\Delta$ Regge trajectories are nearly identical when one plots $M^2$ versus  the  hadron angular momentum $J$.   For example, the spectroscopy of $q \bar q$ mesons are well represented by the simple form $M^2(n,L) = 4\kappa^2(n+ L + {S\over 2})$ where $L$ is the internal relative orbital angular momentum.  The baryons obey a similar formula.  The slopes of the mesons  and baryon trajectories are identical, not  just in $L$,  but also in the principal quantum number $n$  -- despite the fact that mesons are 
$q \bar q$  bound states and baryons are bound states of three quarks.  In fact,  as seen in Fig.~\ref{NSTARFigC}, all of the Regge trajectories for mesons and baryons  consisting of the $u, d$ and $s$ quarks have a universal slope, both in angular momentum $M^2 \propto 4\kappa^2 L$ and in the radial quantum number $n$: $M^2 \propto 4 \kappa^2 n$.

\begin{figure}
 \begin{center}
\includegraphics[height=12cm,width=16cm]{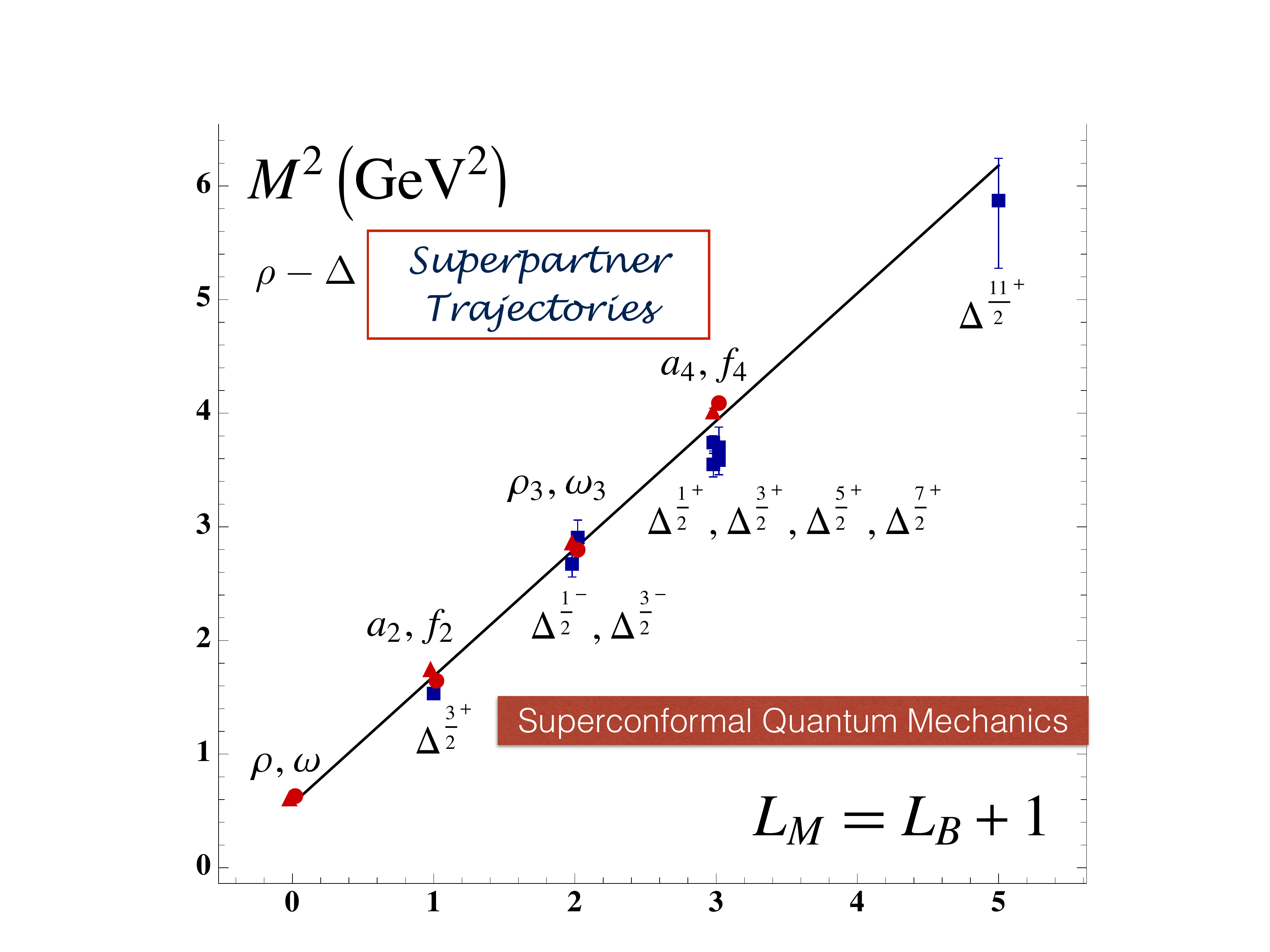}
\label{NSTARFigB}
\end{center}
\caption{Comparison of the $\rho/\omega$ meson Regge trajectory with the $J=3/2$ $\Delta$  baryon trajectory.   
Superconformal algebra  predicts the mass  degeneracy of the meson and baryon trajectories if one identifies a meson with internal orbital angular momentum $L_M$ 
with its superpartner baryon with $L_M = L_B+1.$
See Refs.~\cite{deTeramond:2014asa,Dosch:2015nwa}. }
\end{figure}

The universality of the Regge slopes for both mesons and baryons has been recognized as a fundamental feature of hadron spectroscopy since the early days of hadron physics. 
It  is consistent with the ansatz that the baryons are bound states of a color-$3_C$ quark bound to a  $\bar 3_C$ $qq$ diquark cluster, with the same color-confining dynamics which binds a $3_C$ quark to a $\bar 3_C$ antiquark in a meson.  The universality of the Regge slopes also indicates that QCD has a universal mass scale $\kappa$ which controls hadron spectroscopy even in the limit of zero quark mass.
In fact,  as seen in Fig. \ref{NSTARFigB} the $\rho$ and $\Delta$ masses nearly coincide when one compares the meson trajectory,  plotted as
$M_M^2 $ versus $L_M$, where $L_M$ is the relative orbital angular momentum between the quark and antiquark,  with the $\Delta$  baryon trajectory,   plotted as $M_B^2$ versus $ L_B$, 
the relative orbital angular momentum between the quark and spin-1 diquark.  The masses of the mesons and baryons then  match if one identifies  $L_M=L_B+1$.    The meson and baryon partners not only have the same mass, but also the same twist $\tau = 2+L_M = 3+ L_B$. This also implies that their form factors have same power-law fall-off at high $Q^2$:
$ F_H(Q^2) \propto {1/ Q^{2(\tau -1)} }$.

The degeneracy between meson and baryon masses and their Regge slopes indicates that QCD has a hidden supersymmetry where the fermionic and bosonic eigensolutions  have the same mass. This property reflects the fact that chiral QCD (with  massless quarks)  is conformally invariant at the semi-classical level.
The conformal group in fact has an elegant $ 2\times 2$ Pauli matrix representation called {\it superconformal algebra}, which was originally discovered by  Haag, Lopuszanski, and Sohnius~\cite{Haag:1974qh}.
For example, the conformal Hamiltonian operator and the special conformal operators can be represented as anticommutators of Pauli matrices
$H = {1/2}[Q, Q^\dagger]$ and  $K = {1/2}[S, S^\dagger]$.  

\begin{figure}
 \begin{center}
\includegraphics[height=12cm,width=16cm]{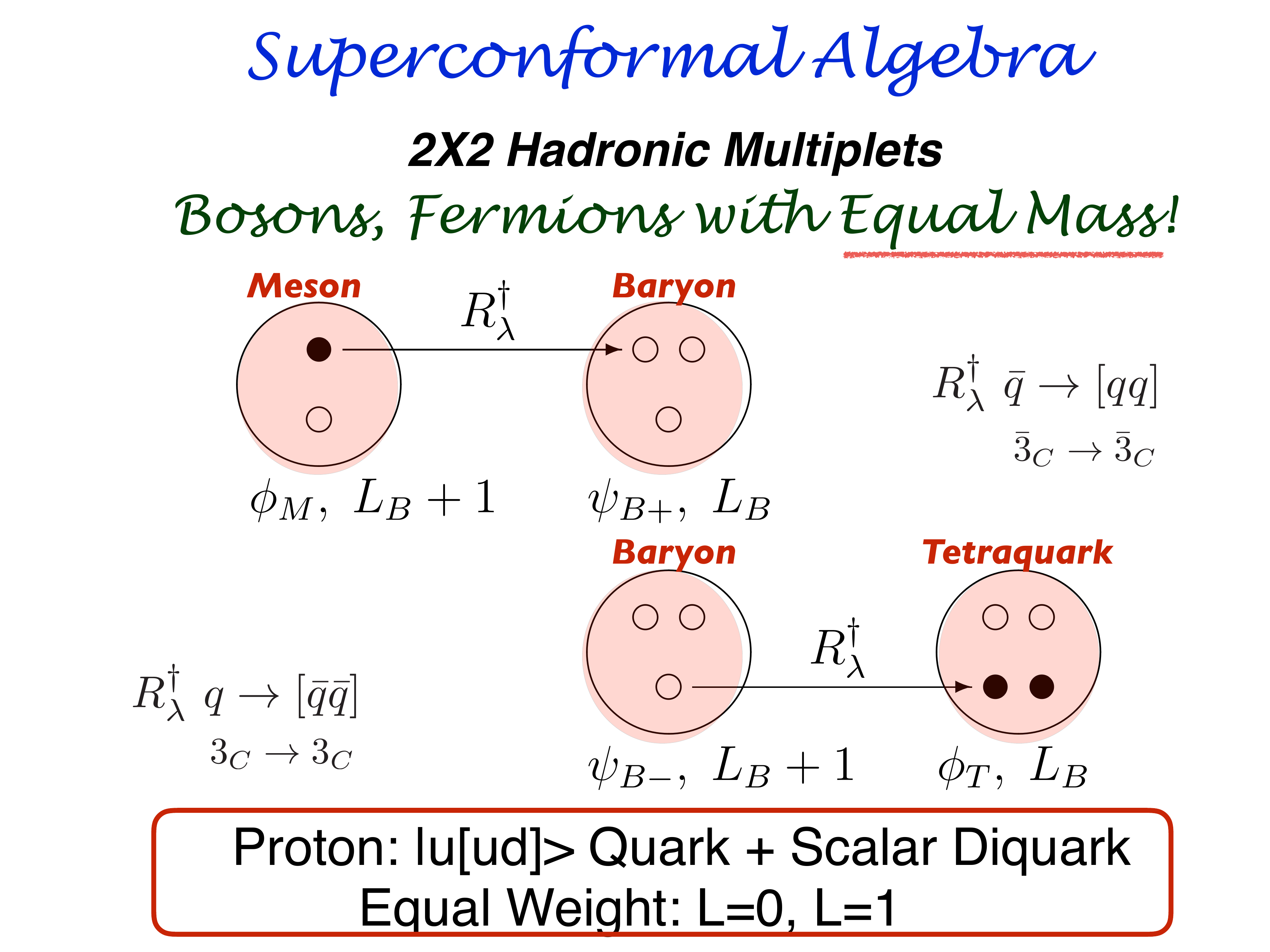}
\label{NSTARFigD}
\end{center}
\caption{The 4-plet representation of mass-degenerate hadronic states predicted by superconformal algebra~\cite{Brodsky:2013ar}. Mesons are $ q \bar q$ bound states, baryons are quark plus anti-diquark bound states and  tetraquarks are diquark plus antidiquark bound states. The supersymmetric ladder operator $R^\dagger_\lambda $ connects quarks and  anti-diquark clusters of the same color. The baryons have two Fock states with orbital angular momentum $L_B$ and $L_B+1$ with equal weight. 
The predicted  meson baryon and tetraquark masses  are identical if one identifies a meson with internal orbital angular momentum $L_M$ with its superpartner baryon or tetraquark with $L_B = L_M-1$.}
\end{figure}

The mass degeneracy between mesons and baryons can be interpreted in terms of the fundamental 4-plet representation of superconformal algebra.   The hadronic entries of the 4-plet  are illustrated  in Fig. \ref{NSTARFigD}.  Mesons are $ q \bar q$ bound states, and baryons are quark plus anti-diquark bound states.    Each baryon eigenstate has two entries of equal weight corresponding to Fock states with relative orbital angular momentum $L_B$ and $L_B+1$.  In the case of the nucleon, the quark spin $S^z_q=\pm 1/2$ has equal weight to  be parallel or antiparallel to the baryon spin $J^z=\pm 1/2$.   In fact, two Fock states with different $L$ are needed in order for a baryon to have a nonzero anomalous magnetic moment, a nonzero Pauli form factor~\cite{Brodsky:1980zm}, as well as to generate the Sivers single-spin asymmetry~\cite{Brodsky:2002cx} in deep inelastic lepton-nucleon scattering.  The  four-plet  also includes tetraquarks -- bound states of diquarks and anti-diquarks -- with the same mass as their meson and baryonic partners.   

The supersymmetric ladder operator $R^\dagger_\lambda $ connects quarks and  anti-diquark clusters of the same color.  It  connects the baryon and meson states  and  their Regge trajectories to each other in a remarkable manner; in fact, superconformal algebra  predicts that the bosonic meson and fermionic baryon masses are equal if one identifies each meson with internal orbital angular momentum $L_M$ with its superpartner baryon with $L_B = L_M-1$; the meson and baryon  superpartners  then have the same parity.  Since
$ 2+ L_M = 3 + L_B$, the twist-dimension of the meson and baryon superpartners are also the same.   Superconformal algebra thus explains the phenomenological observation that Regge trajectories  for both mesons and baryons have parallel slopes.  As discussed below, this symmetry and its mass degeneracies also  consistent with light-front holography, the duality between light-front quantization and AdS/QCD.

\begin{figure}
\begin{center}
\includegraphics[height=12cm,width=16cm]{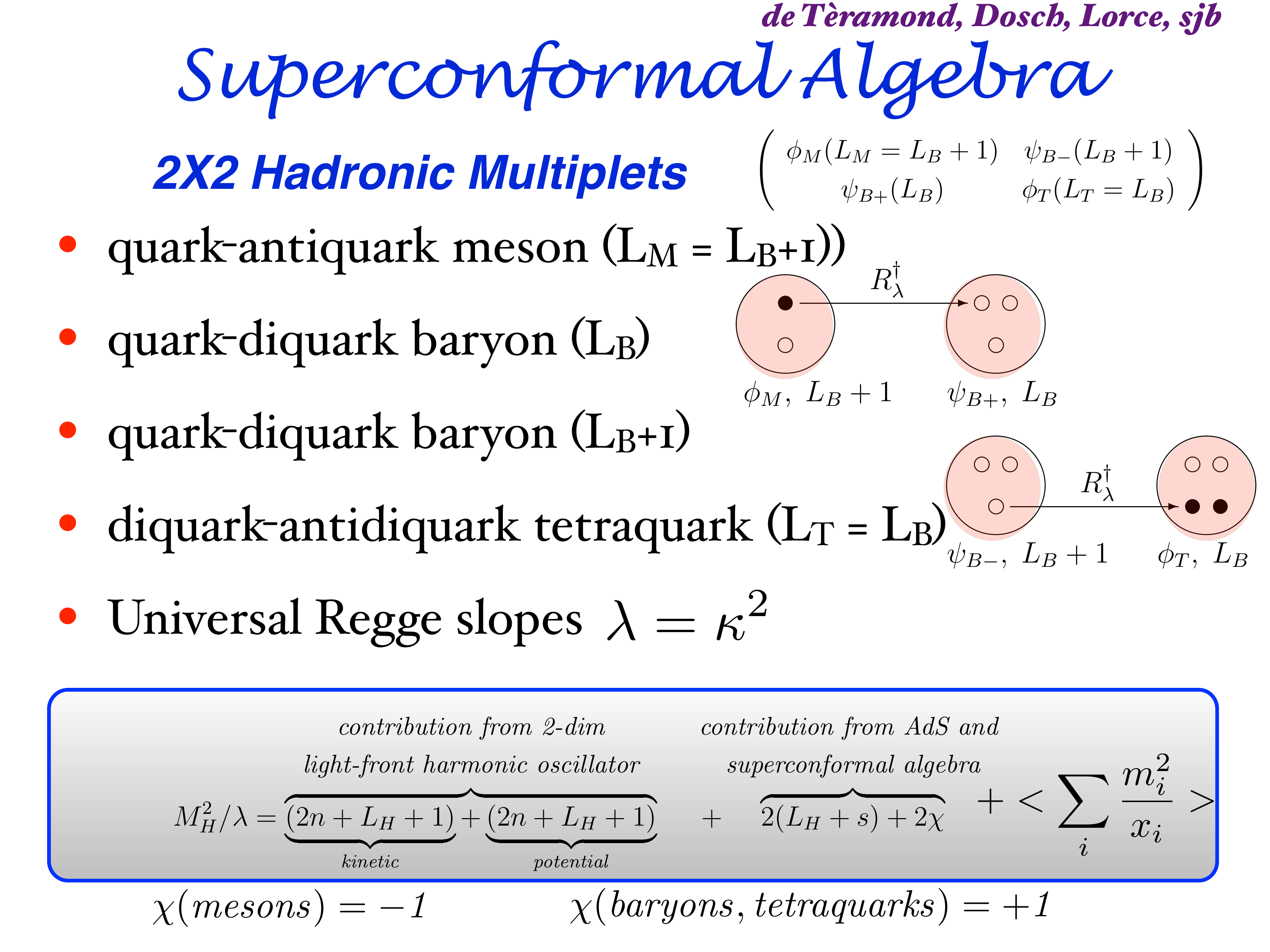}
\end{center}
\caption{The eigenstates of superconformal algebra have a $2 \times 2 $ representation of mass degenerate bosons and fermions:  a meson with $L_M=  L_B+1$, a baryon doublet with 
$L_B, L_B+1$ components and a tetraquark with $ L_T = L_B$. The breakdown of LF kinetic, potential, spin, and quark mass contributions to each hadron is also shown.  The virial theorem predicts the equality of the LF kinetic and potential contributions.
}
\label{2X2Multiplets}
\end{figure} 

Thus, as illustrated in Fig. \ref{2X2Multiplets}, the hadronic eigensolutions of the superconformal algebra are   $2\times 2$ matrices connected internally by the supersymmetric algebra operators.
The eigensolutions of the supersymmetric conformal algebra have a Pauli matrix representation, where the upper-left component corresponds to mesonic $q \bar q$ color-singlet bound states, the two off-diagonal eigensolutions $\psi^{\pm}$  correspond to a pair of Fock components of baryonic quark-diquark states with equal weight, where the quark spin is parallel or antiparallel to the baryon spin, respectively.  The fourth component corresponds to diquark anti-diquark (tetraquark) bound states. 
The resulting frame-independent color-confining bound-state LF eigensolutions can be identified with the hadronic eigenstates of confined quarks  for $SU(3)$ color.    In effect, two of the quarks of the baryonic  color singlet $qqq$ bound state  bind to a color $\overline{3_C}$ diquark bound state, which then binds by the same color force to the remaining $3_C$ quark.  As shown by t'Hooft in a string model~\cite{tHooft:2004doe}, the $Y$ configuration of three quarks is unstable, and it reduces to the quark-diquark configuration.  The matching of the meson and baryon spectra is thus due to the fact that the same color-confining potential that binds two quarks to a diquark  also  binds a quark to an antiquark.

The same slope controls the Regge trajectories of both mesons and baryons in both the orbital angular momentum $L$ and the principal quantum number $n$.
Only one mass parameter $\kappa$  appears; it sets the confinement scale and the hadron mass scale in the  chiral limit, as well as  the length scale which underlies hadron structure.    In addition to the meson and baryon eigenstates, one also predicts color-singlet {\it tetraquark}  diquark-antidiquark bound states with the same mass as the baryon.

\section{Origin of the QCD Mass Scale}

One of the fundamental, profound questions in hadron physics is how a nonzero hadronic mass such as the proton mass can emerge from QCD since there is no explicit  parameter with mass dimensions in the chiral QCD Lagrangian with zero quark mass.    This dilemma is compounded by the fact that  the chiral QCD Lagrangian has no knowledge of the conventions used for the units of mass such as $MeV$.     Thus QCD with $m_q=0$ can in principle only predict {\bf ratios of masses } such as $m_\rho/m_p$ -- not the absolute values. Similarly,  given that color is confined, how does QCD set its range without a parameter with dimensions of length?   It is hard to see how this mass scale problem could be solved 
by `` dimensional transmutation", since a mass parameter  determined by  perturbative QCD such as $\Lambda_{\overline MS}$, is renormalization-scheme dependent, whereas hadron masses  are independent of the conventions chosen to regulate the UV divergences. 

An important principle,  first demonstrated by  de Alfaro, Fubini and Furlan  (dAFF)~\cite{deAlfaro:1976je} for conformal theory in $1+1$ quantum mechanics, is that a fundamental mass scale can appear in a Hamiltonian and its equations of motion without 
affecting the conformal invariance of the action.  The essential step introduced by dAFF is to add to the conformal Hamiltonian terms proportional to the dilation operator $D$ and the special conformal operator $K$. The coefficients have mass units with arbitrary values.   The consequence of this linear transformation is the addition of a harmonic oscillator potential $V(x) = \kappa^4 x^2$ to the Hamlitonian and thus confinement.    The algebra of the conformal group is effectively  maintained.  In fact, the new Hamitonian has ``extended dilatation invariance"  since the mass scale $\kappa$ can be rescaled arbitrarily.  The mass scale $\kappa$ is never determined in absolute units -- thus only ratios of the mass eigenvalues can be determined, not their absolute values.  dAFF then show that one can redefine the time variable $t \to \tau$ such that the action retains its conformal invariance,  The new time variable $\tau$ has support only over a finite interval. However, a  finite range of $\tau$ is consistent with the fact that the interval in time measured between confined constituents is always finite.

As shown by Fubini and Rabinovici,~\cite{Fubini:1984hf},  a nonconformal Hamiltonian with a mass scale and universal confinement can also be obtained for superconformal algebra by shifting $Q \to Q +\omega K$, the analog of the dAFF procedure. The conformal algebra can be extended even though $\omega$ has dimension of mass.  
In effect, one generates generalized supercharges of the superconformal algebra~\cite{Fubini:1984hf}.   The result of this shift of the Hamiltonian is a again a color-confining harmonic  potential in the equations of motion, and remarkably the action remains conformally invariant; again, only one mass parameter appears.

\begin{figure}
 \begin{center}
\includegraphics[height=12cm,width=16cm]{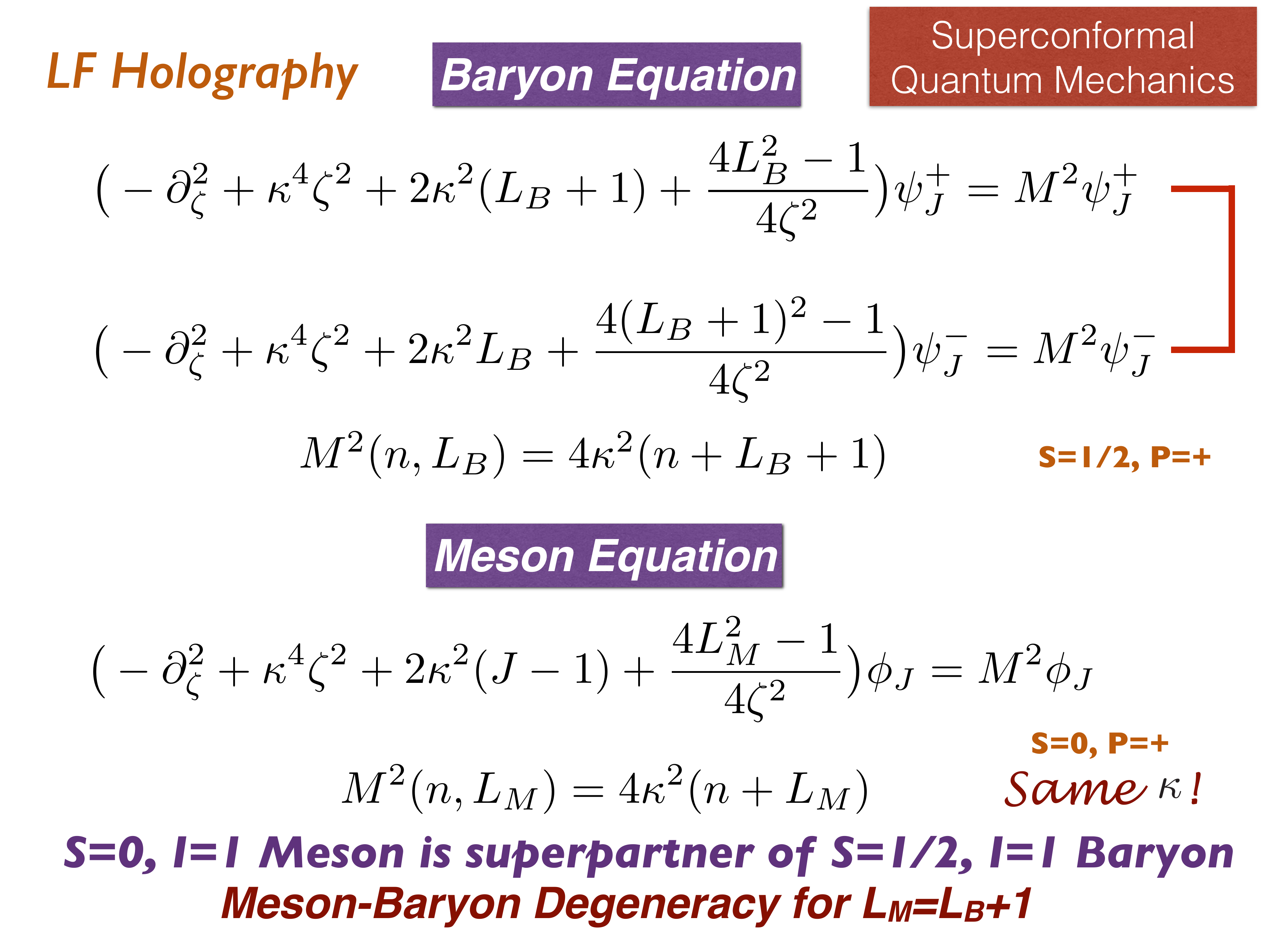}
\label{NSTARFigJ}
\end{center}
\caption{The LF Schr\"odinger equations for baryons and mesons for zero quark mass derived from the Pauli $2\times 2$ 4-plet matrix representation of superconformal algebra.  
The $\psi^\pm$  are the baryon quark-diquark LFWFs where the quark spin $S^z_q=\pm 1/2$ is parallel or antiparallel to the baryon spin $J^z=\pm 1/2$.   The predicted meson and baryon masses are identical if one identifies a meson with internal orbital angular momentum $L_M$ with its superpartner baryon with $L_B = L_M-1.$
See Refs.~\cite{deTeramond:2014asa,Dosch:2015nwa,Dosch:2015bca}.}
\end{figure}

De T\'eramond, Dosch, and I~\cite{Brodsky:2013ar}
have shown that a mass gap and color confinement appears when one extends the dAFF procedure to relativistic, causal, Poincar\'e invariant, light-front Hamiltonian theory for QCD. Light-front quantization at fixed light-front time $\tau=t+z/c$  provides a physical, frame-independent formalism for hadron dynamics and structure.    The bound-state equations of superconformal algebra are, in fact, Lorentz-invariant, frame-independent, relativistic light-front Schrodinger equations.   
The equations for zero mass quarks are shown in Fig. \ref{NSTARFigJ}. 
The resulting eigensolutions are thus eigenstates of a supersymmetric light-front Hamiltonian obtained from $AdS_5$  and light-front holography.  
The predictions for both hadronic spectroscopy and dynamics provide an elegant description of meson and baryon phenomenology, including Regge trajectories with universal slopes in the principal quantum number $n$ and the orbital angular momentum $L$.  In addition, the resulting quark-antiquark meson bound-state equation predicts that the  pion eigenstate with $n=L=S=0$ is massless for zero quark mass.

Superconformal algebra leads to effective QCD light-front bound-state equations for both mesons and baryons~\cite{deTeramond:2014asa,Dosch:2015nwa,Dosch:2015bca}. 
The resulting set of bound-state equations for confined quarks are shown in Fig. \ref{NSTARFigJ}.   The same equation for mesons is also obtained from light-front holography using AdS$_5$ modified by the dilaton $e^{+\kappa^2 z^2}.$   
The confinement potential in the LF formalism has the form $U(\zeta^2) = \kappa^4 \zeta^2$ for light quarks.   Here $\zeta^2$  is the LF radial variable conjugate to the $q \bar q$ invariant mass. It yields the familiar $\sigma r $ potential for heavy quark $Q \bar Q$ quarkonium states in the nonrelativistic limit ~\cite{Trawinski:2014msa}.

\begin{figure}
\begin{center}
\includegraphics[height=16cm,width=18cm]{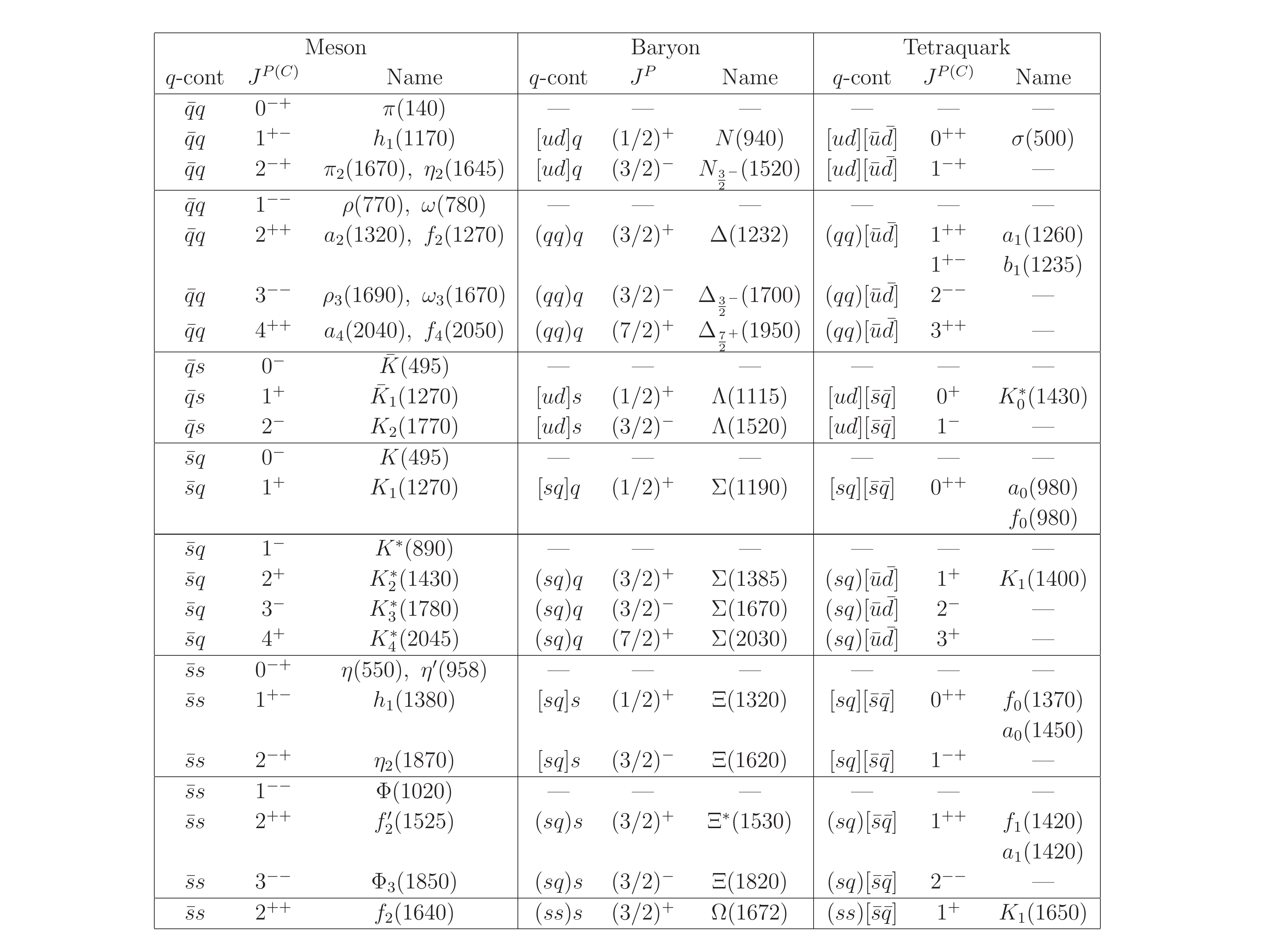}
\end{center}
\caption{Classification and quantum numbers of mesons, baryons, and tetraquarks composed of light quarks related by superconformal algebra\cite{Nielsen}.  }
\label{Nielsen.pdf}
\end{figure}

A comparison of the meson and baryon masses of the $\rho/\omega$ Regge trajectory with the spin-$3/2$ $\Delta$ trajectory 
is shown in Fig. \ref{NSTARFigB}.  The observed hadronic spectrum  with $N_C=3$  are seen to exhibit the supersymmetric features predicted by superconformal algebra.   
One thus obtains a unified Regge spectroscopy of meson, baryon, and tetraquarks, including remarkable supersymmetric relations between the masses of mesons and baryons and a universal Regge slope.    See Fig. \ref{Nielsen.pdf}~\cite{Nielsen}.  A similar classification can be applied to light-heavy hadrons such as the $D$ and $B$ mesons.   A detailed comparison of hadron spectroscopy with the predicted tetraquark spectrum will be given in ref. \cite{Nielsen}.

\begin{figure}
\begin{center}
\includegraphics[height=12cm,width=16cm]{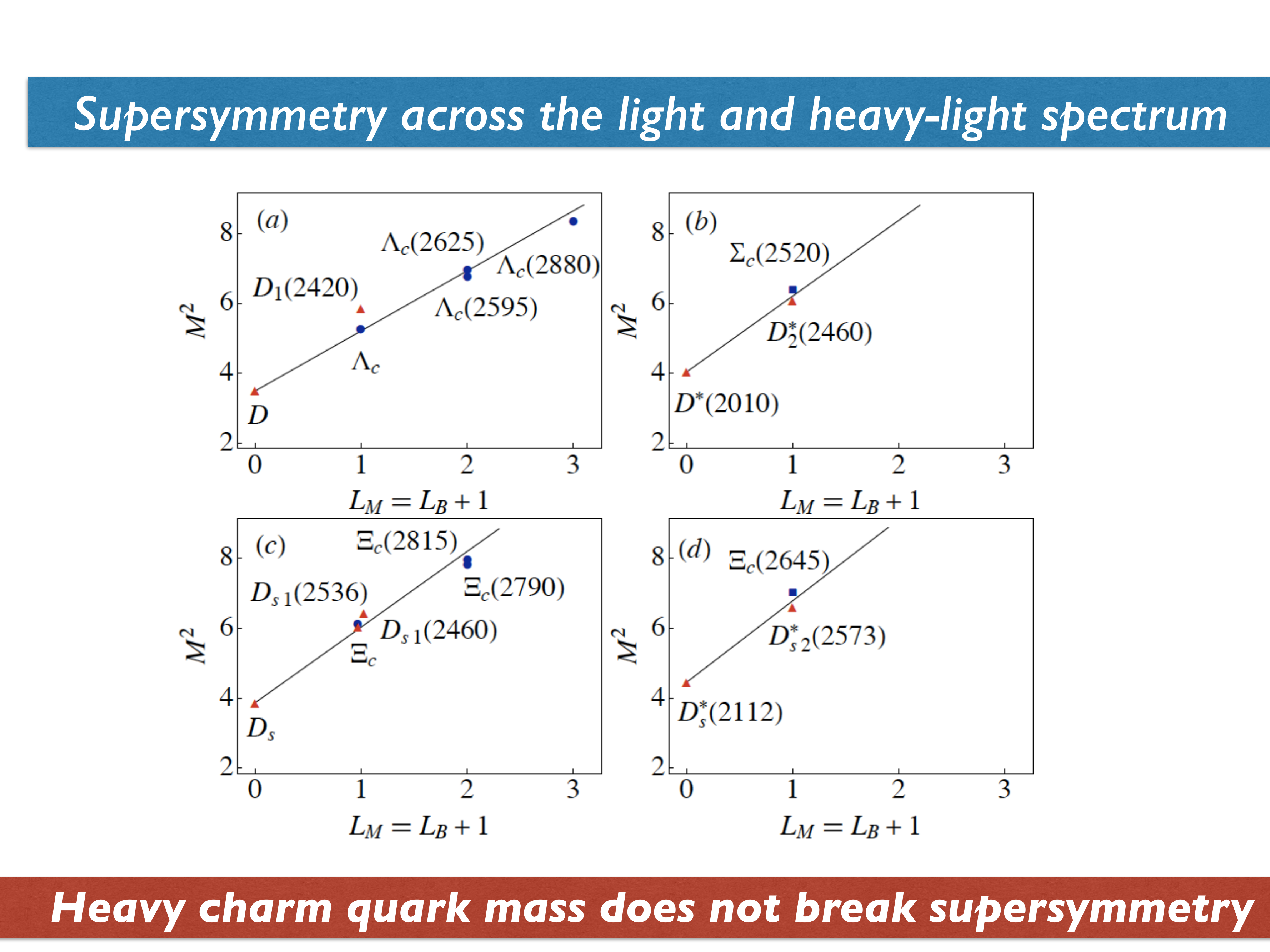}
\end{center}
\caption{Comparison of the meson and baryon Regge trajectories for hadrons with a single charm quark.}
See Refs.~\cite{deTeramond:2014asa,Dosch:2015nwa}. 
\label{NSTARFigI}
\end{figure}

One can generalize these results to heavy-light $[\bar Q q] $ mesons and  $[Q [qq]]$ baryons~\cite{Dosch:2016zdv}.  Linear Regge trajectories and meson-baryon degeneracy are observed. The Regge slopes are found to increase for heavy quark masses $m_Q$ as expected from heavy-quark effective field theory;  however, the supersymmetric connections between the meson and baryon heavy-light hadrons are maintained. See Fig. \ref{NSTARFigI}.

\begin{figure}
\begin{center}
\includegraphics[height=12cm,width=16cm]{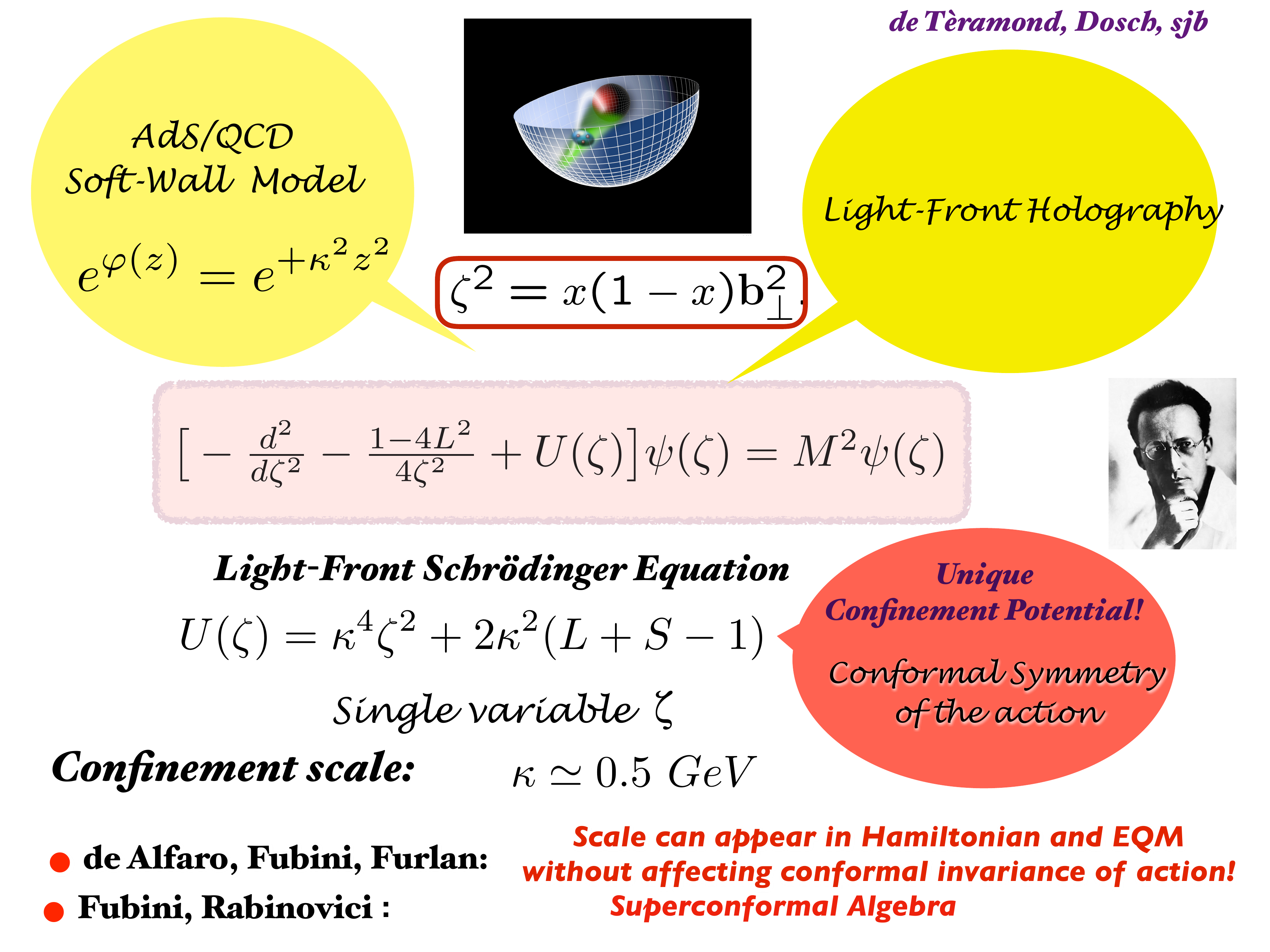}
\end{center}
\caption{The convergence of  theoretical methods for generating a model of hadron spectroscopy and dynamics with color confinement  and meson-baryon supersymmetric  relations.}. 
\label{NSTARFigK}
\end{figure}

The combination of light-front holography with superconformal algebra thus leads to the novel prediction that hadron physics has supersymmetric properties in both spectroscopy and dynamics. As discussed below, it also predicts the analytic form of the QCD running coupling in the nonperturbative domain,  and it provides new insights into the physics underlying hadronization at the amplitude level.  Other advances in holographic QCD and  superconformal algebra are reviewed in refs.~\cite{Brodsky:2016vig,Brodsky:2016nsn,Brodsky:2017tyf}.  The synthesis of AdS/QCD with superconformal algebra and the dAFF ansatz is illustrated in Fig. \ref{NSTARFigK}.

\section{Light-Front QCD}

Light-Front quantization is the natural formalism for relativistic quantum field theory.  Measurements of hadron structure, such as deep inelastic lepton-proton scattering, are made at a fixed light-front time $\tau= t+z/c$, along of the front of a light-wave, in analogy to a flash photograph -- not at a single ``instant time".  As shown by Dirac~\cite{Dirac:1949cp}, boosts are kinematical in the ``front form".  Thus all formulae using the front form are independent of the 
observer's motion~\cite{Brodsky:1997de}; i.e., they are  Poincar\'e invariant.

One can derive the light-front Hamlitonian $H_{LF}$ directly from the QCD Lagrangian and avoid ghosts and longitudinal  gluonic degrees of freedom by choosing the light-cone gauge  $A^+ =0$.  
Quark masses appear in the LF kinetic energy as $\sum_i {m^2\over x_i}$. This can be derived from the Higgs theory quantized using LF dynamics~\cite{Srivastava:2002mw}.
The confined quark field $\psi_q$ couples to the background Higgs field  $g_{\overline \Psi_q } <H >  \Psi_q$ via its Yukawa  scalar matrix element  coupling  
$g_q <H > \bar u(p) 1 u(p) = m_q \times {m_q \over x} = {m^2\over x}.$  The usual  Higgs vacuum expectation value  $<H > $ is replaced by a constant zero mode when one quantizes the Standard Model using light-front quantization~\cite{Srivastava:2002mw}.

The eigenstates of the light-front (LF) Hamiltonian $ H_{LF} =  P^+ P^- -{\vec P}^2_\perp$ derived from the QCD Lagrangian encodes the entire the hadronic mass spectrum for both individual hadrons and  the multi-hadron continuum.   The eigenvalues of the LF Hamiltonian  are the squares of the hadron masses $M^2_H$: 
$H_{LF}|\Psi_H>  = M^2_H |\Psi_H>$~\cite{Brodsky:1997de}.   The evaluation of the Wilson line for gauge theories in the front form is discussed in ref.~\cite{Reinhardt:2016fjl}.
There are also advantages for perturbative QCD calculations using light-front-time-ordered perturbation theory, including the use of $J^z$ conservation.

Here  $P^- = i {d\over d\tau}$ is the LF time evolution operator, and $P^+=P^0+P^z$ and $\vec P_\perp$ are kinematical.  
The eigenfunctions of $H_{LF} $ provide the hadronic LF Fock state wavefunctions (LFWFs):
 $ \psi^H_n(x_i, \vec k_{\perp i },\lambda_i)= <n| \Psi_H> $, the projection of the hadronic eigenstate on the free Fock basis. The constituents' physical momenta are 
$p^+_i = x_i P^+$, and  $\vec p_{\perp i } =  x_i  {\vec P}_\perp +  \vec k_{\perp i }$,  and the $\lambda_i$ label the  spin projections $S^z_i$.  Remarkably, one can reduce the LF Hamiltonian theory for mesons with $m_q=0$   to an effective  LF Schrodinger equation for the valence $q \bar q$ Fock state in  terms of a single variable --  the LF radial variable 
$\zeta^2 = b^2_\perp x(1-x)$. The same equation is obtained using LF holography and suoerconformal algebra.

The LFWFs are Poincar\'e invariant: they are independent of $P^+$ and $P_\perp$ and are thus independent of the motion of the observer.  Since the LFWFs are independent of the hadron's momentum, there is no physical effects analogous to ``length contraction"~\cite{Terrell:1959zz,Penrose:1959vz}. Structure functions are computed from the absolute square of the frame-independent LFWFs.
One thus measures the  same structure function in an electron-proton collider as in the traditional deep inelastic electron-proton scattering experiment where the target nucleon is at rest.

\begin{figure}
 \begin{center}
\includegraphics[height= 14cm,width=16cm]{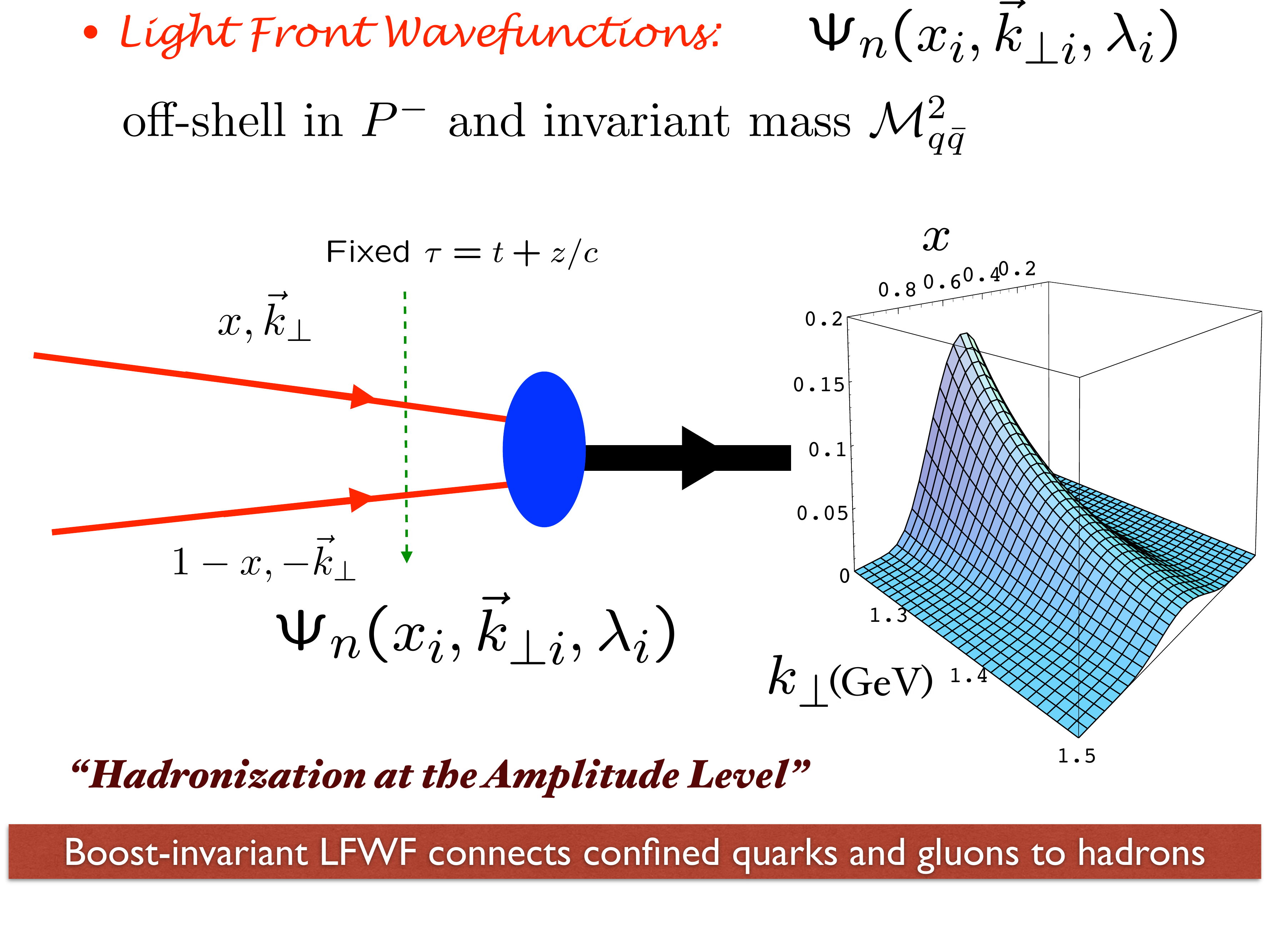}
\end{center}
\caption{ The meson LFWF connects the intermediate $q \bar q $ state, which is off of the $P^-$ energy shell and thus off-the-invariant mass shell  ${\cal M}^2  > m^2_H$T to the  physical meson state with 
${\cal M}^2  = m^2_H$.    The $q$ and $\bar q$ can be regarded as effective dressed fields}. 
\label{had1}
\end{figure}

Light-front wavefunctions, the  eigensolutions  of the  QCD light-front Hamiltonian $H_{LF}^{QCD}$  thus provide a direct link between the fundamental QCD Lagrangian and hadron structure. Since they are defined at a fixed $\tau$, they connect the physical on-shell hadronic state to its quark and gluonic constituents, not at off-shell energy, but off-shell in $P^-$ and thus, equivalently, the invariant mass squared ${\cal M}^2 =( \sum_i k^\mu_i )^2.$ 
The LF wavefunctions thus also control the transformation of quarks and gluons in an off-shell intermediate state into an observed final on-shell hadronic state; i.e., hadronization at the amplitude level~\cite{Brodsky:2009dr}.   For example, as illustrated in Fig.~\ref{had1},  the meson LFWF connects  the off-the-invariant mass shell quark and antiquark to the on-shell asymptotic physical meson state. The QED analog of atom formation is discussed in Ref.~\cite{Munger:1993kq}.

LFWFs thus play the same role in hadron physics as the Schr\"odinger wavefunctions which encode the structure of atoms in QED.  The elastic and transition form factors of hadrons, weak-decay amplitudes and distribution amplitudes are overlaps of LFWFs; structure functions, transverse momentum distributions
and other inclusive observables 
are constructed from the squares of the LFWFs.     In contrast, one cannot compute the form factors of hadrons or other current matrix elements from the overlap of the usual fixed-time $t$ ``instant" form wavefunctions since one must also include contributions  where the photon interacts with connected, but acausal, vacuum-induced currents.
The calculation of deeply virtual Compton scattering using LFWFs is given in Ref.~\cite{Brodsky:2000xy}.  One can also compute the gravitational form factors of hadrons.  In particular, 
one can show that the anomalous gravitomagnetic moment $B(q^2=0)$ vanishes identically for any LF Fock state~\cite{Brodsky:2000ii}, in agreement with the equivalence theorem of gravity~\cite{Kobzarev:1962wt,Teryaev:1999su}.

\subsection{Solving  LF Hamiltonian Theory}

The LF Heisenberg equation can in principle be solved numerically by matrix diagonalization  using the ``Discretized Light-Cone  Quantization" (DLCQ)~\cite{Pauli:1985pv} method.  Anti-periodic boundary conditions in 
$x^-$ render the $k^+$ momenta  discrete  as well as  limiting the size of the Fock basis.   In fact, one can easily solve $1+1 $ quantum field theories such as QCD$(1+1)$~\cite{Hornbostel:1988fb} for any number of colors, flavors and quark masses using DLCQ. 
Unlike lattice gauge theory, the nonpertubative DLCQ analysis is in Minkowski space, is frame-independent, and is free of fermion-doubling problems.   
AdS/QCD, based on the  AdS$_5$ representation of the conformal group in five dimensions, maps to physical 3+1 space-time at fixed LF time;  this correspondence, ``light-front holography"~\cite{deTeramond:2008ht},  is  provides a  color-confining approach  to $H_{LF}^{QCD}$ for QCD(3+1).  An improved method for solving nonperturbative QCD,  ``Basis Light-Front Quantization" (BLFQ)~\cite{Vary:2014tqa},  uses the eigensolutions of a color-confining approximation to QCD (such as LF holography) as the basis functions,  rather than the plane-wave basis used in DLCQ, thus incorporating the full dynamics of QCD.  LFWFs can also be determined from the covariant 
Bethe-Salpeter wavefunction by integrating over $k^-$~\cite{Brodsky:2015aia}.  
A review of the light-front formalism is given in Ref.~\cite{Brodsky:1997de}.

The full  QCD LF equation for massless quarks can be reduced to an  effective LF Shr\"odinger radial equation for the valence  $|q \bar q> $ Fock state of $q \bar q$ mesons
 $$\large[-{d^2\over d\zeta^2} + {4 L^2-1\over 4 \zeta^2 } + U(\zeta^2) \large] \psi = M^2 \psi$$ 
 and similar bound-state equations for baryons, represented as  quark + diquark-cluster $ |q [qq]>$  eigenstates.  Only one variable $\zeta^2$ appears after projecting on a Fock state with fixed $L^z$. 
The  ``radial"  LF variable $\zeta^2= b^2_\perp x(1-x)$ of LF theory is conjugate  to the LF kinetic energy.   The identical equation is derived from $AdS_5$, where the fifth coordinate $z$ is identified with $\zeta$ (Light Front Holography). 

The color-confining potential $U(\zeta^2) = \kappa^4 \zeta^2 + 2\kappa^2(J-1) $ can be derived from soft-wall $AdS_5$,  incorporating the remarkable dAFF principle that a mass scale can appear in the Hamiltonian while retaining the conformal invariance of the action.  
The result is a color-confining LF potential which depends on a single universal constant $\kappa$  with mass dimensions.   In addition, by utilizing superconformal algebra~\cite{Dosch:2015bca}, the resulting hadronic color-singlet eigenstates have a $2 \times 2 $ representation of mass-degenerate bosons and fermions:   a $|q \bar q>$ meson with $L_M=  L_B+1$, a baryon doublet $|q [qq]>$ with 
$L_B$ and  $L_B+1$ components of equal weight,  and a tetraquark $|[qq] [\bar q \bar q] >$ with $ L_T = L_B$.  See: Fig. \ref{2X2Multiplets}.   Thus ratios of hadron masses such as $m_\rho = {M_p\over \sqrt 2}$ are predicted. The individual contributions  LF kinetic energy, potential energy, spin-interactions, and the quark mass to the mass squared of each hadron is also shown.  The virial theorem for harmonic oscillator confinement predicts the equality of the LF kinetic and LF potential contributions to $M^2_H$ for each hadron.

.

\subsection{LF Perturbation Theory}

LF-time-ordered perturbation theory  can be advantageous for computing perturbative QCD amplitudes..  
An example  of LF-time-ordered perturbation theory is the computation of multi-gluon scattering amplitudes by Cruz-Santiago and Stasto~\cite{Cruz-Santiago:2015dla}.  
In this method, the propagating particles  are all on their respective mass shells:  $k_\mu k^\mu = m^2$, and intermediate states are off-shell in invariant mass;  {\it i.e.}, $P^- \ne \sum k^-_i$.  Unlike instant form, where one must sum  $n!$ frame-dependent  amplitudes, only $\tau$-ordered diagrams where each propagating particle has  positive $k^+ =k^0+k^z$  can contribute. 
The number of nonzero amplitudes is also greatly reduced by noting that the total angular momentum projection $J^z = \sum_i^{n-1 } L^z_i + \sum^n_i S^z_i$ and the total $P^+$ are  conserved at each vertex~\cite{Chiu:2017ycx}. 

A remarkable advantage of LF time-ordered perturbation theory (LFPth) is that the calculation of a subgraph of any order in pQCD only needs to be done once;  the result can be stored in a ``history" file.  This is due to the fact that in LFPth the numerator algebra is independent of the process; the denominator changes, but only by a simple shift of the initial $P^-$.   Another simplification is that loop integrations are three dimensional: $\int d^2\vec k_\perp \int^1_0 dx.$   Unitarity is  explicit and 
renormalization can be implemented using the ``alternate denominator" method which defines the required subtraction counterterms~\cite{Brodsky:1973kb}.

A key property of light-front quantization is $J^z$ conservation~\cite{Chiu:2017ycx}; the $z$-component of  angular momentum remains unchanged under Lorentz transformations generated by the light-front kinematical boost operators.   
Particles in the front form move with positive $k^+ = k^0+k^z \ge 0.$ 
The quantization axis for $J^z$ for each particle is the same axis $\hat z$ which defines LF time $\tau = t + z/c$.  The spin along
the $\hat z$ direction defined by the light-front Lorentz transformation is preserved because $<J^3>_{LF}= S^z$ for all momenta $p^\mu$.  $J^z$ conservation underlies the Jaffe spin sum rule~\cite{Jaffe:1987sx}. 
Thus $S^z$ and $L^z$ refer to angular momentum in the $\hat z$ direction. 
As in nonrelativistic quantum mechanics,  $J^z =  \sum^n_{i=1} S^z_i + \sum^{n-1}_{n=1}  L^z_i$ for any $n$- particle intermediate or Fock state.  There are $n-1$ relative orbital angular momenta.   It is conserved at every vertex and is conserved overall for any process and ``LF helicity" refers to the spin projection $S^z$ of each particle and ``LF chirality"  is the spin projection $S^z$  for massless particles.   In a renormalizable theory $L^z$ can only change  by one unit at any vertex.  This leads to a rigorous selection rule for amplitudes at fixed order in pQCD~\cite{Chiu:2017ycx}: $|\Delta L^z |  \le n$ in an
$n$-th order perturbative expansion.  This  selection rule for the orbital angular momentum  can be used to eliminate interaction vertices in QED and QCD,  and it provides an upper bound on the change of orbital angular momentum in scattering processes at any fixed order in perturbation theory.

By definition, spin and helicity can be used interchangeably in the front form.  
LF chirality is conserved by the vector current in electrodynamics and the axial current of electroweak interactions.   Each coupling  conserves quark chirality when the quark mass  is  set to zero.    A compilation of LF spinor matrix elements is given in Ref.~\cite{Lepage:1980fj}.

 Light-front spin, quantized in the $\hat z$ direction, is not the same  as the usual ``Wick helicity",  where helicity is defined as the projection of the spin on the particle's three-momentum $\vec k$.  
Wick helicity is thus not conserved unless all particles move in the same direction.  Wick helicity can be frame dependent. 
For example, In the case of  $gg \to H$, the  Wick helicity assignment  is $(+1) + (+1) \to 0$ in the CM frame,  but it is 
$(+1) + (-1) \to 0$  for collinear gluons if the  two gluons move in the same direction.

The twist of a hadronic interpolating operator corresponds to the number of fields plus the total $|L^z|$.   The  pion LF Fock state  for  $\pi \to q \bar q$ with  twist-2 corresponds to $(J^z_\pi=0) \to (S^z_q =+ {1\over 2} ) +  (S^z= - {1\over 2} )$ with zero relative orbital angular momentum $L^z_{q \bar q}$.   
This is the Fock state of the pion that decays to $\ell \nu$ via the LF chiral-conserving axial current $\gamma^\mu \gamma_5$. 
The twist-3 pion in the OPE corresponds to 
$J^z_\pi =0 \to (S^z_q =+ {1\over 2} ) +  (S^z_{\bar q}= + {1\over 2} ) + (L^z =-1) $  or $ J^z=0 \to (S^z_q = - {1\over 2} ) +  (S^z_{\bar q}= - {1\over 2} ) + (L^z_{q \bar q} =-1), $
where $L^z$ is the relative orbital angular momentum between the quark and antiquark.  The twist-3 Fock state couples the pion to the chiral-flip pseudoscalar $\gamma_5$  operator.
The GMOR relation connects the twist-2 and twist-3 Fock states when $m_q \ne 0$~\cite{Brodsky:2012ku}. 
The twist-3 proton  with $J^z_p=+{1\over 2}$ in AdS/QCD is a bound state of a quark  with $S^z_p={1\over 2}$ and a spin-zero diquark $[qq]$  with $L^z_{q [qq]} = 0$, and the twist-4 proton in AdS/QCD is a bound state of a quark with $S^z_p=-{1\over 2}$ and spin-zero diquark $[qq]$  with relative orbital angular momentum $L^z_{q [qq]} = +1)$.  LF holography predicts equal probability for the twist-3 and twist-4 Fock states in the nucleon for $m_q=0.$

\section{Light-Front Holography} 
 
\begin{figure}
 \begin{center}
\includegraphics[height= 14cm,width=16cm]{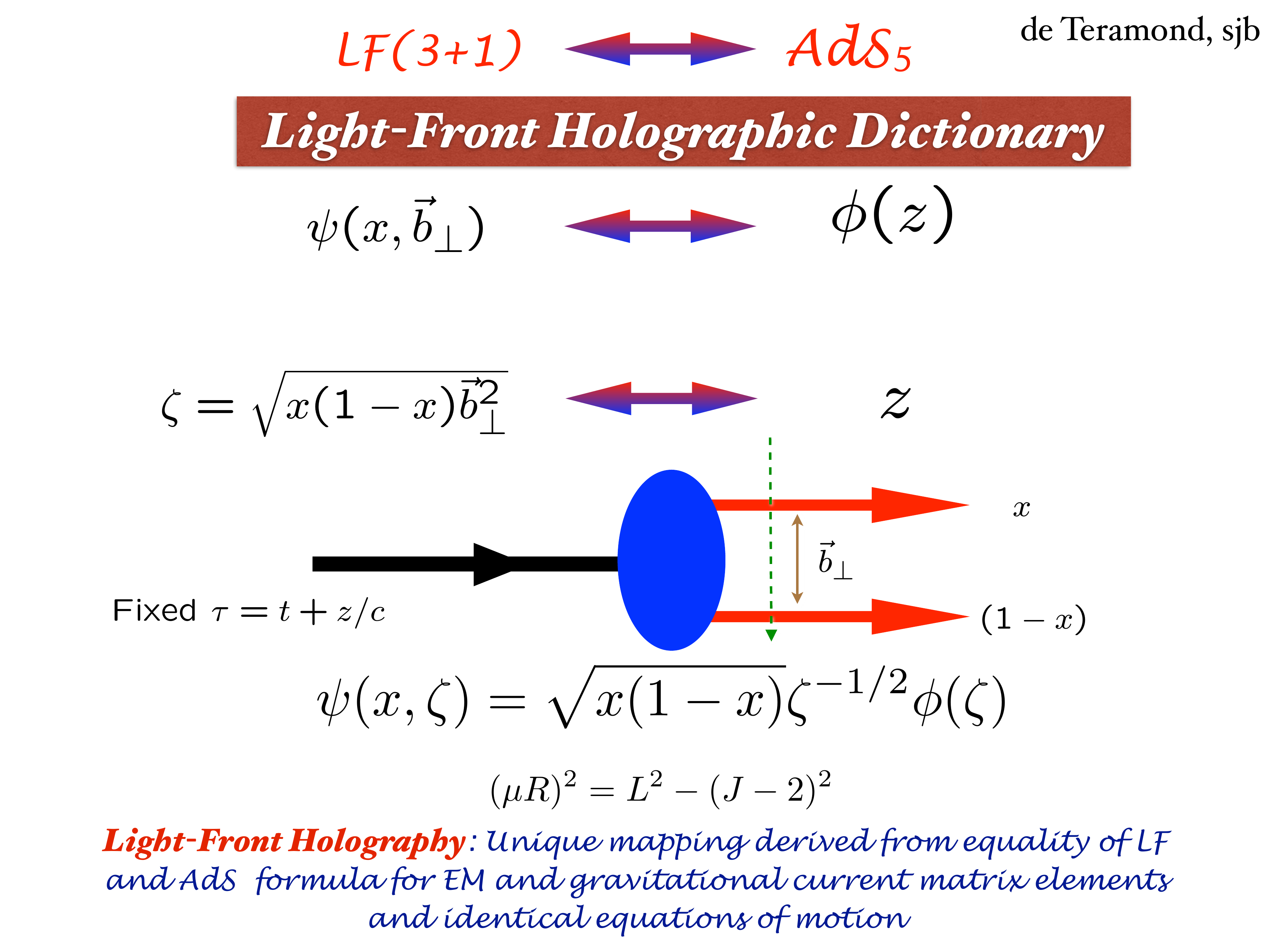}
\end{center}
\caption{The holographic dictionary which maps the fifth dimension variable $z$ of  the five-dimensional AdS$_5$ space to the LF radial variable $\zeta$ where 
$\zeta^2  =  b^2_\perp(1-x)$.   The same physics transformation maps the AdS$_5$  and $(3+1)$ LF expressions for electromagnetic and gravitational form factors to each other. 
From Ref.~\cite{deTeramond:2013it}}
\label{dictionary}
\end{figure}

Five-dimensional AdS$_5$ space provides a geometrical representation of the conformal group.
The color-confining light-front  equation for mesons of arbitrary spin $J$ can be derived~\cite{deTeramond:2013it}
from the holographic mapping of  the ``soft-wall model" modification of AdS$_5$ space for the specific dilaton profile $e^{+\kappa^2 z^2},$  where one identifies the fifth dimension coordinate $z$ with the light-front coordinate $\zeta$. Remarkably ,  AdS$_5$  is holographically dual to $3+1$  spacetime at fixed light-front time $\tau = t+ z/c$.  
The holographic dictionary is summarized in Fig.~\ref{dictionary}   An important feature of AdS/QCD is that hyperfine spin terms that appear in the effective LF Hamiltonian which split the $\pi$ and $\rho$ masses, etc.,  are automatically predicted.

The combination of light-front dynamics, its holographic mapping to AdS$_5$ space, and the dAFF procedure provides new insight into the physics underlying color confinement, the nonperturbative QCD coupling, and the QCD mass scale.  A comprehensive review is given in  Ref.~\cite{Brodsky:2014yha}.  The $q \bar q$ mesons and their valence LF wavefunctions are the eigensolutions of the frame-independent relativistic bound state LF Schr\"odinger equation -- the same meson equation that is derived using superconformal algebra.
The mesonic $q\bar  q$ bound-state eigenvalues for massless quarks are $M^2(n, L, S) = 4\kappa^2(n+L +S/2)$.
The equation predicts that the pion eigenstate  $n=L=S=0$ is massless at zero quark mass. The  Regge spectra of the pseudoscalar $S=0$  and vector $S=1$  mesons  are 
predicted correctly, with equal slope in the principal quantum number $n$ and the internal orbital angular momentum $L$.  A comparison with experiment is shown in Fig. \ref{ReggePlot.pdf}.

The LF Schr\"odinger Equations for baryons and mesons derived from superconformal algebra  are shown  in Fig. 6.
As explained above, the baryons on the proton (Delta) trajectory are bound states of a quark with color $3_C$ and scalar (vector)  diquark with color $\bar 3_C$ 
The proton eigenstate labeled $\psi^+$ (parallel quark and baryon spins) and $\psi^-$ (anti parallel quark and baryon spins)  have equal Fock state probability -- a  feature of ``quark chirality invariance".  Since the nucleon Fock states with $S^z_q = \pm 1/2$ have equal weight, all of the nucleon spin is carried by quark orbital angular momentum. Predictions for the static properties of the nucleons are discussed in ref.~\cite{Liu:2015jna}.

Superconformal algebra also predicts that the LFWFs of the superpartners are related, and thus the corresponding elastic and transition form factors are identical.   The resulting  predictions for meson and baryon timelike form factors can be tested in $e^+ e^- \to H \bar H^\prime $ reactions.

\begin{figure}
\begin{center}
\includegraphics[height=12cm,width=16cm]{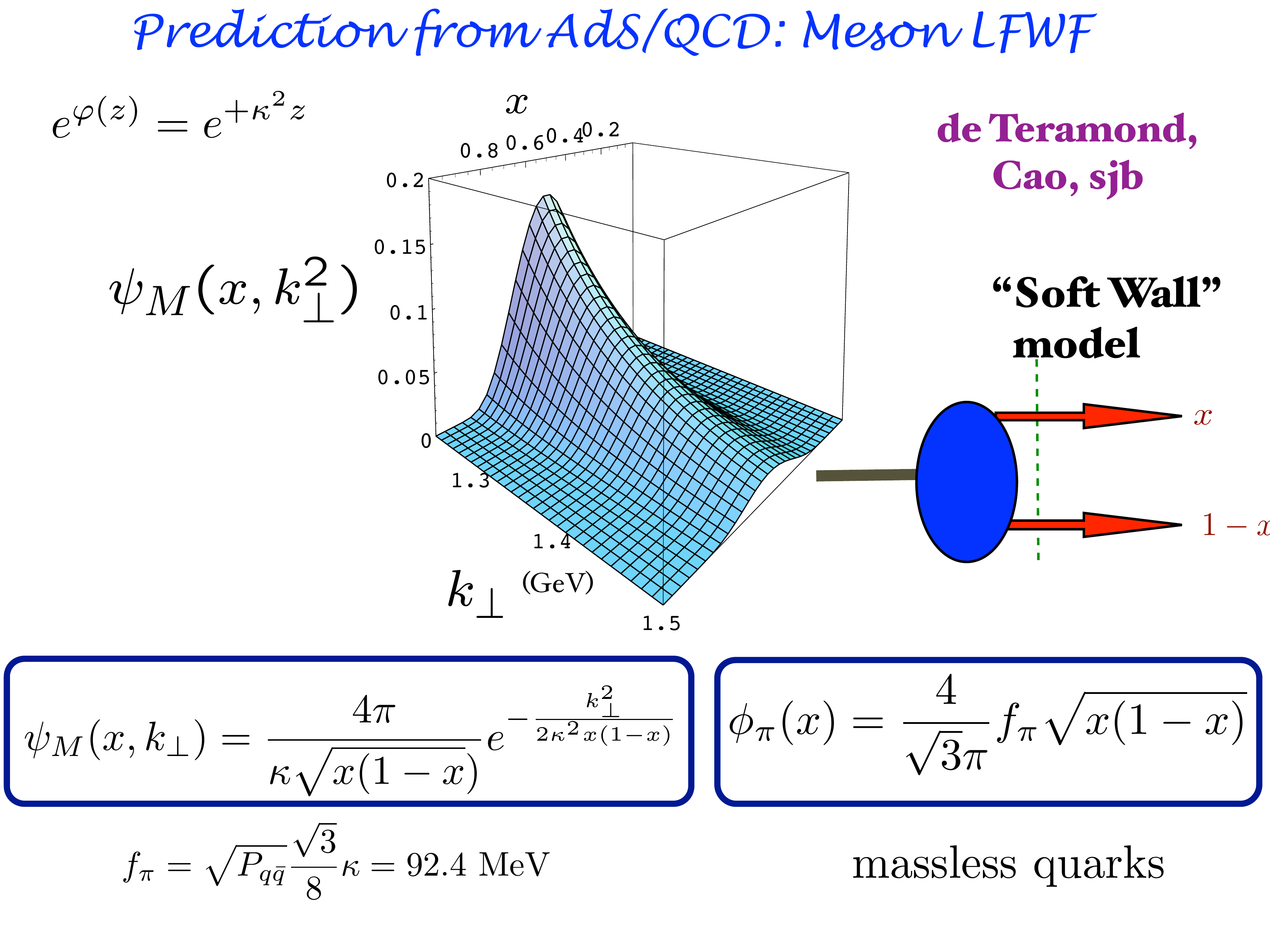}
\end{center}
\caption{Prediction from AdS/QCD and Light-Front Holography for  meson LFWFs  $\psi_M(x,\vec k_\perp)$   and the pion distribution amplitude.     
}
\label{MesonLFWF.pdf}
\end{figure}

\begin{figure}
 \begin{center}
\includegraphics[height= 14cm,width=16cm]{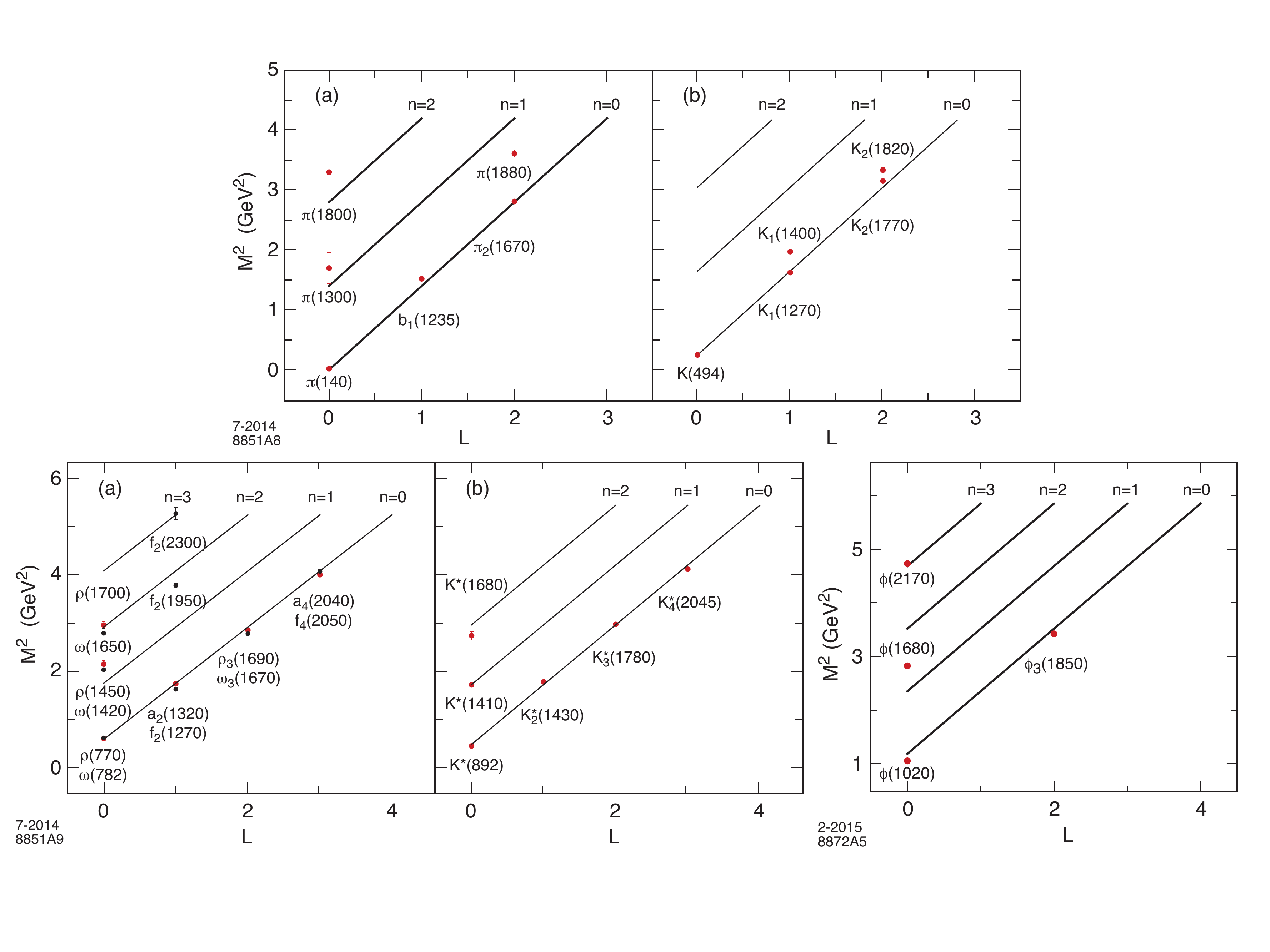}
\end{center}
\caption{Comparison of the AdS/QCD prediction  $M^2(n, L, S) = 4\kappa^2(n+L +S/2)$ for the orbital $L$ and radial $n$ excitations of the meson spectrum with experiment.   The pion is predicted to be massless for zero quark mass. The $u,d,s$ quark masses can be taken into account by perturbing in $<m_q^2/x>$.   The fitted value of $\kappa = 0.59$ GeV for pseudoscalar mesons, 
and  $\kappa = 0.54$ GeV  for vector mesons. }
\label{ReggePlot.pdf}
\end{figure}

The hadronic LFWFs predicted by light-front holography and superconformal algebra  are  functions of the LF kinetic energy $\vec k^2_\perp/ x(1-x)$ -- the conjugate of the LF radial variable $\zeta^2 = b^2_\perp x(1-x)$ -- times a function of $x(1-x)$; they do not factorize as a  function of $\vec k^2_\perp$ times a function of $x$.  The resulting  nonperturbative pion distribution amplitude $\phi_\pi(x) = \int d^2 \vec k_\perp \psi_\pi(x,\vec k_\perp) = (4/  \sqrt 3 \pi)  f_\pi \sqrt{x(1-x)}$,  see Fig.~\ref{MesonLFWF.pdf}, which controls hard exclusive process, is  consistent with the Belle data for the photon-to-pion transition form factor~\cite{Brodsky:2011xx}.  
The AdS/QCD light-front holographic eigenfunction for the $\rho$ meson LFWF $\psi_\rho(x,\vec k_\perp)$ gives excellent 
predictions for the observed features of diffractive $\rho$ electroproduction $\gamma^* p \to \rho  p^\prime$,  as shown by Forshaw and Sandapen~\cite{Forshaw:2012im}

The contribution to the mass squared of the hadrons from LF kinetic energy, LF potential energy and spin interactions is also shown in Fig.~\ref{2X2Multiplets}.
The first-order contribution $<{m^2_q\over x}> $ from nonzero quark masses; i.e. the coupling to the  background Higgs zero mode is also indicated.  The equality of the LF kinetic energy and LF potential energy reflects the virial theorem for harmonic confinement $\kappa^4 \zeta^2.$

\begin{figure}
 \begin{center}
\includegraphics[height= 14cm,width=16cm]{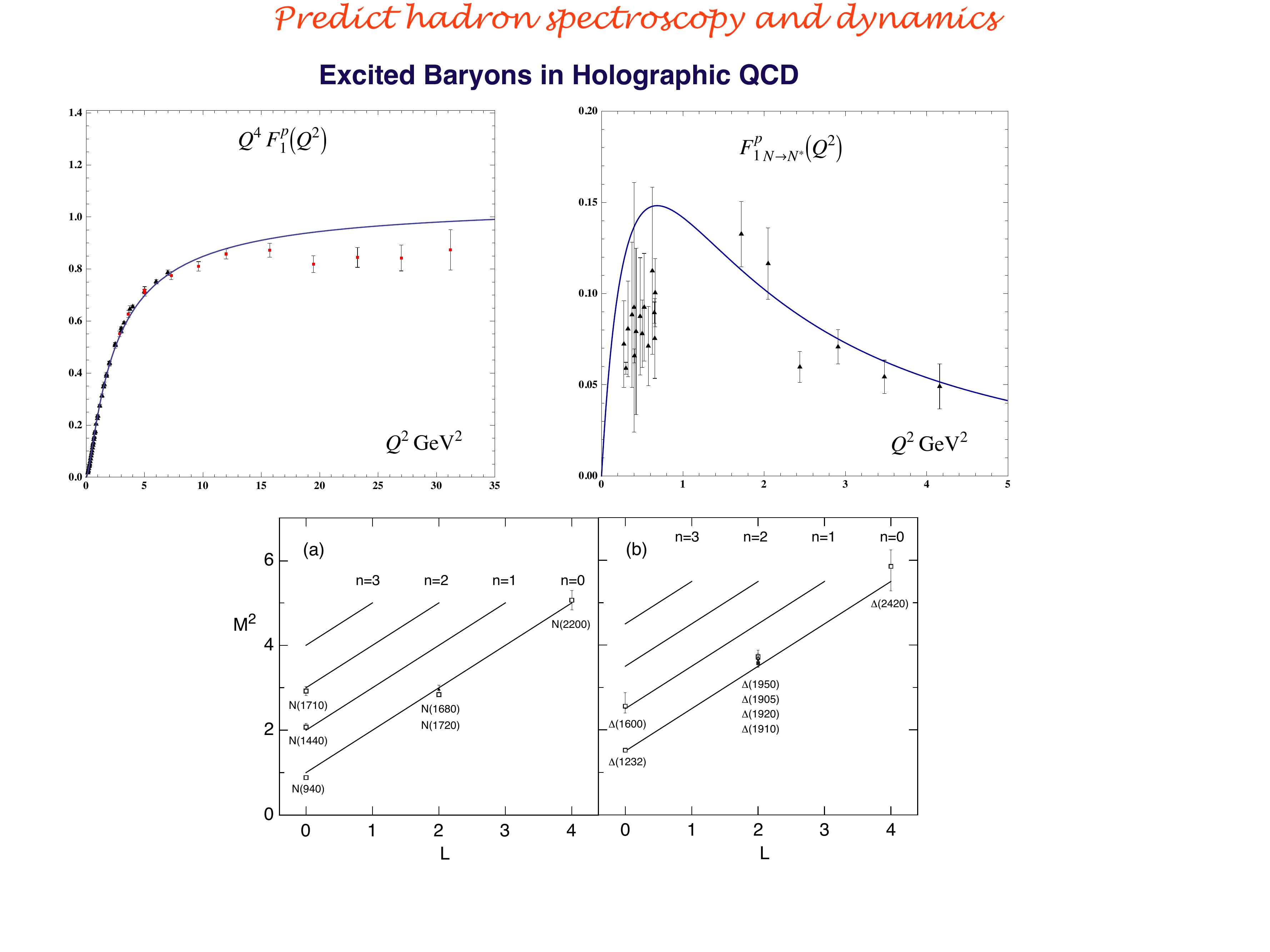}
\end{center}
\caption{Predictions for baryon elastic and transition spacelike form factors using light-front holographic QCD (LFHQCD) and superconformal algebra. }
\label{NSTARFigF}
\end{figure}

\begin{figure}
 \begin{center}
\includegraphics[height= 14cm,width=16cm]{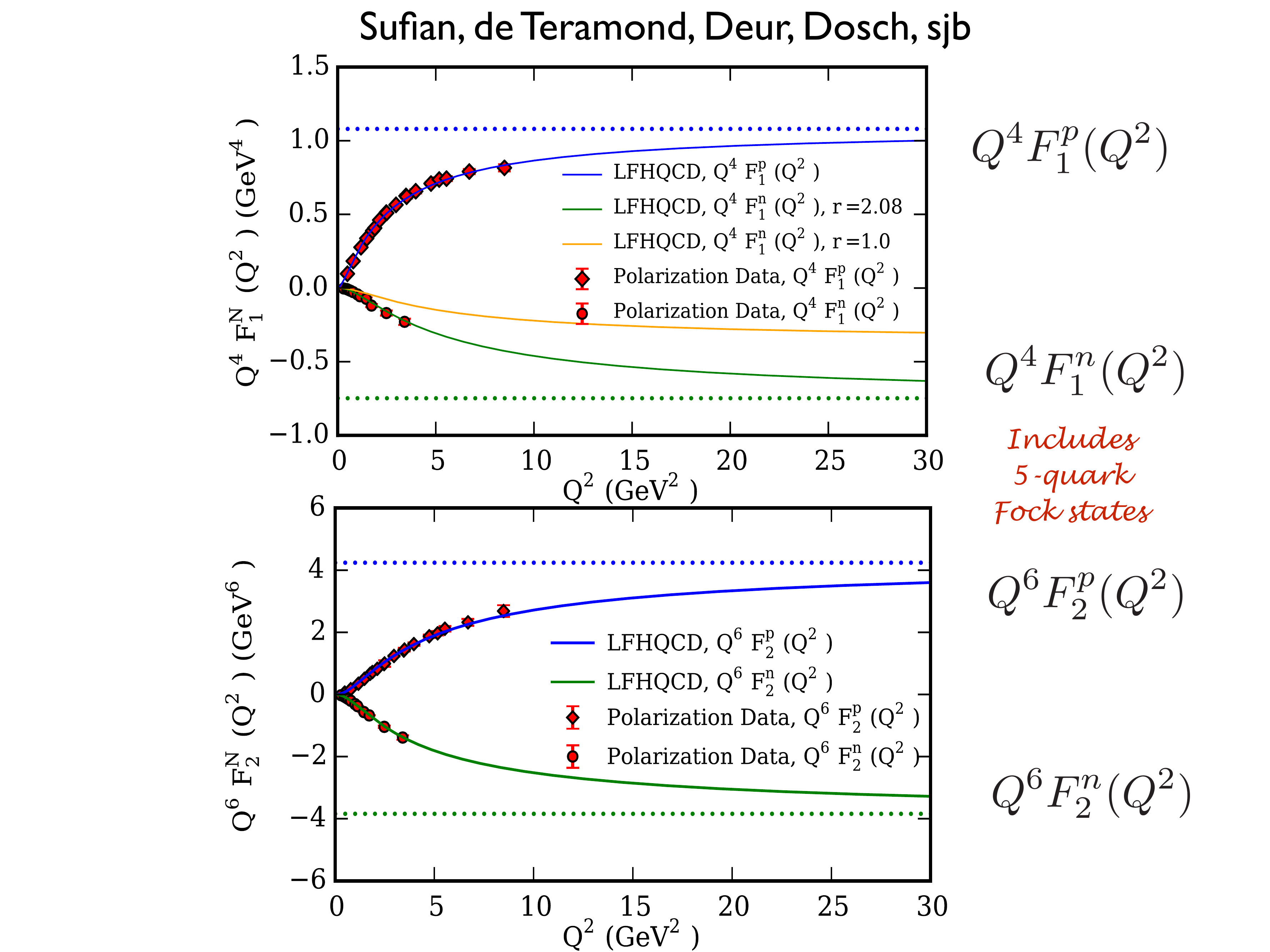}
\end{center}
\caption{Predictions for baryon elastic and transition spacelike form factors using LFHQCD and superconformal algebra.  The fit allows for  five-quark $|uud q \bar q >$ Fock state  in the proton.}
\label{NSTARFigG}
\end{figure}

The combination of Light-Front Holography and Superconformal Algebra not only predicts meson and baryon  spectroscopy  successfully, but also hadron dynamics, such as  vector meson electroproduction,  hadronic light-front wavefunctions, distribution amplitudes, form factors, and valence structure functions.   

Since the LF wavefunctions are determined one can also predict the form factors of each hadron.  For example, the proton form factors and baryon transition form factors can be predicted from the overlap of LF wavefunctions using the Drell-Yan-West formula. Predictions are shown in Fig. \ref{NSTARFigF}. Counting rules are automatically maintained.  
One can also augment the predictions using a small percentage of 5-quark Fock states as motivated by a meson cloud.  The predictions for this model are shown in Fig. \ref{NSTARFigG}. In addition, one can predict timelike form factors such as that measured in $e^+ e^- \to \pi^+ \pi^-$ reactions. The vector meson poles appear in the timelike amplitudes when one uses the dressed current predicted by AdS/QCD. 
An application to the deuteron elastic form factors and structure functions is given 
in ref.~\cite{Gutsche:2015xva,Gutsche:2016lrz}

\subsection{Color Confinement from LF Holography}

Remarkably, the light-front potential using the dAFF procedure has the unique form of a harmonic oscillator $\kappa^4 \zeta^2$ in the 
light-front invariant impact variable $\zeta$ where $ \zeta^2Ê = b^2_\perp x(1-x)$. The result is  a single-variable frame-independent relativistic equation of motion for  $q \bar q $ bound states, a ``Light-Front Schr\"odinger Equation"~\cite{deTeramond:2008ht}, analogous to the nonrelativistic radial Schr\"odinger equation in quantum mechanics. This result, including spin terms, is obtained using  light-front holography  -- the duality between the front form and AdS$_5$, the space of isometries of the conformal group -- if one  
modifies the action of AdS$_5$ by the dilaton $e^{\kappa^2 z^2}$ in the fifth dimension $z$.  The  Light-Front Schr\"odinger Equation  incorporates color confinement and other essential spectroscopic and dynamical features of hadron physics, including a massless pion for zero quark mass and linear Regge trajectories with the identical slope  in the radial quantum number $n$   and internal  orbital angular momentum $L$.      
As shown above when one generalizes this procedure using superconformal algebra, the resulting light-front eigensolutions predict a unified Regge spectroscopy of meson, baryon, and tetraquarks, including remarkable supersymmetric relations between the masses of mesons and baryons of the same parity.

It is interesting to note that the contribution of the {\it `H'} diagram to $Q \bar Q $ scattering is IR divergent as the transverse separation between the $Q$  
and the $\bar Q$ increases~\cite{Smirnov:2009fh}.  This is a signal that pQCD is inconsistent without color confinement.  The sum of such diagrams could sum to the confinement potential $\kappa^4 \zeta^2 $, as dictated by the dAFF principle that the action remains conformally invariant, despite the appearance of the mass scale $\kappa$ in the Hamiltonian.
The $\kappa^4 \zeta^2$ confinement interaction between a $q$ and $\bar q$ will also induce a $\kappa^4/s^2$ correction to $R_{e^+ e^-}$, replacing the $1/ s^2$ signal usually attributed to a vacuum gluon condensate.

It should be emphasized that the value of the mass scale $\kappa$ is not determined in an absolute sense by QCD.    Only ratios of masses are determined, and the theory has an effective dilation or scale invariance under $\kappa \to C \kappa $.    In a sense, chiral QCD has an ``extended conformal invariance."  The resulting time variable  which retains the conformal invariance of the action has finite support, conforming to the fact that the LF time between the interactions with the confined constituents is finite.  

\subsection{Positronium-Proton scattering}

\begin{figure}
 \begin{center}
\includegraphics[height= 10cm,width=16cm]{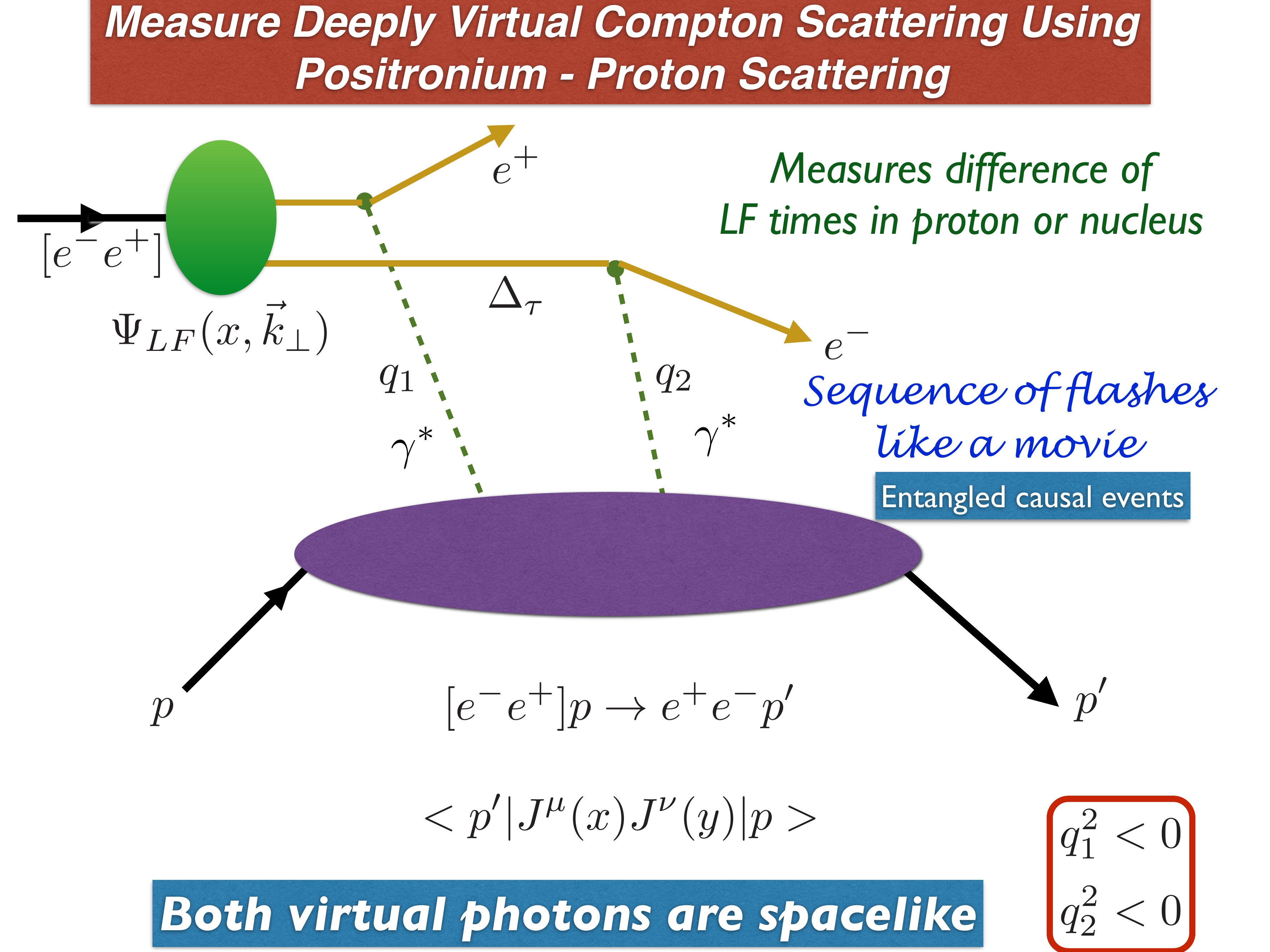}
\end{center}
\caption{Doubly Virtual Compton scattering on a proton (or nucleus) can be measured for two {\it spacelike} photons $q^2_1, q^2_2 <0$ 
with minimal, tunable, skewness $\xi$ using positronium-proton scattering $[e^+ e^-] p \to e^+ e^- p'$.  
One can also measure double deep inelastic scattering and elastic positronium-proton scattering.  
One can also create a beam of  ``true muonium" atoms $[\mu^- \mu^-]$~\cite{Brodsky:2009gx,Banburski:2012tk} using Bethe-Heitler pair production just below threshold.
 }
\label{Positronium}
\end{figure}

The finite time difference $\Delta \tau$ between the LF times  $\tau$ of the quark constituents of the proton could be measured using positronium-proton scattering $[e^+ e^-] p \to e^+ e^- p'$.  This process, which measures double diffractive deeply virtual Compton scattering for two spacelike photons, is illustrated in Fig.~\ref{Positronium}.  One can produce a relativistic positronium beam  using the collisions of laser photons with high energy photons or by 
using Bethe-Heitler pair production below the $e^+ e^-$ threshold.
The production of parapositronium via the collision of photons is analogous to pion production in two-photon interactions and Higgs production via gluon-gluon fusion.  

One can also measure the LFWFs of QED atoms using diffractive dissociation.

For example,  it is possible to study the dissociation of relativistic positronium
atoms to an electron and positron with light front momentum fractions $x$ and $1-x$ and  opposite transverse momenta.
The LFWF of positronium is the central input.
For example, suppose one creates a relativistic positronium beam.   It will dissociate by Coulomb exchange in a thin target:
$[e^+ e^- ] + Z \to e^+ e^- Z$.
The momentum distribution of the leptons in  the LF variables $x$ and $k_\perp$ will determine   the first derivative of the atomic LFWF 
${d\over d k_\perp} \psi(x,\vec k_\perp)$.
When ${k^2_\perp \over x(1-x)} > 4 m^2_e$ one can observe the transition from NR Schr\"odinger theory  where $ \psi(x,\vec k_\perp) \propto {1\over k^4_\perp}$
to the relativistic domain, where $ \psi(x,\vec k_\perp) \propto {1\over k^2_\perp}$.
One can also test predictions computed from BLFQ (Basis Light-Front Quantization)~\cite{Vary:2013kma}.  Higher Fock states are also possible, such as  $[e^+ e^-] + Z \to e^+ e^-  \gamma Z$ and 
$[e^+ e^-] + Z \to e^+ e^-  e^+ e^- Z$.

Positronium dissociation is  analogous to the E791 Ashery measurements of the pion LFWF at FermiLab
$\pi A \to Jet Jet A$ ~\cite{Ashery:2005wa}, where one observes the transition from Gaussian fall-off to power law  fall-off at large $1\over k^2_\perp $ as predicted by AdS/QCD.
Similarly, one could also measure the LFWF of a nucleus like a deuteron by dissociating relativistic ions $d A \to p n A$ .   At large $1\over k^2_\perp $ one  could observe the transition to the ``hidden-color" Fock states predicted by QCD~\cite{Brodsky:1983vf}.

\section {The QCD Coupling at all Scales} 

The QCD running coupling $\alpha_s(Q^2)$
sets the strength of  the interactions of quarks and gluons as a function of the momentum transfer $Q$.
The dependence of the coupling
$Q^2$ is needed to describe hadronic interactions at 
both long and short distances. 
The QCD running coupling can be defined~\cite{Grunberg:1980ja} at all momentum scales from a perturbatively calculable observable, such as the coupling $\alpha^s_{g_1}(Q^2)$, which is defined from measurements of the Bjorken sum rule.   At high momentum transfer, such ``effective charges"  satisfy asymptotic freedom, obey the usual pQCD renormalization group equations, and can be related to each other without scale ambiguity by commensurate scale relations~\cite{Brodsky:1994eh}.  

\begin{figure}
\begin{center}
\includegraphics[height=12cm,width=16cm]{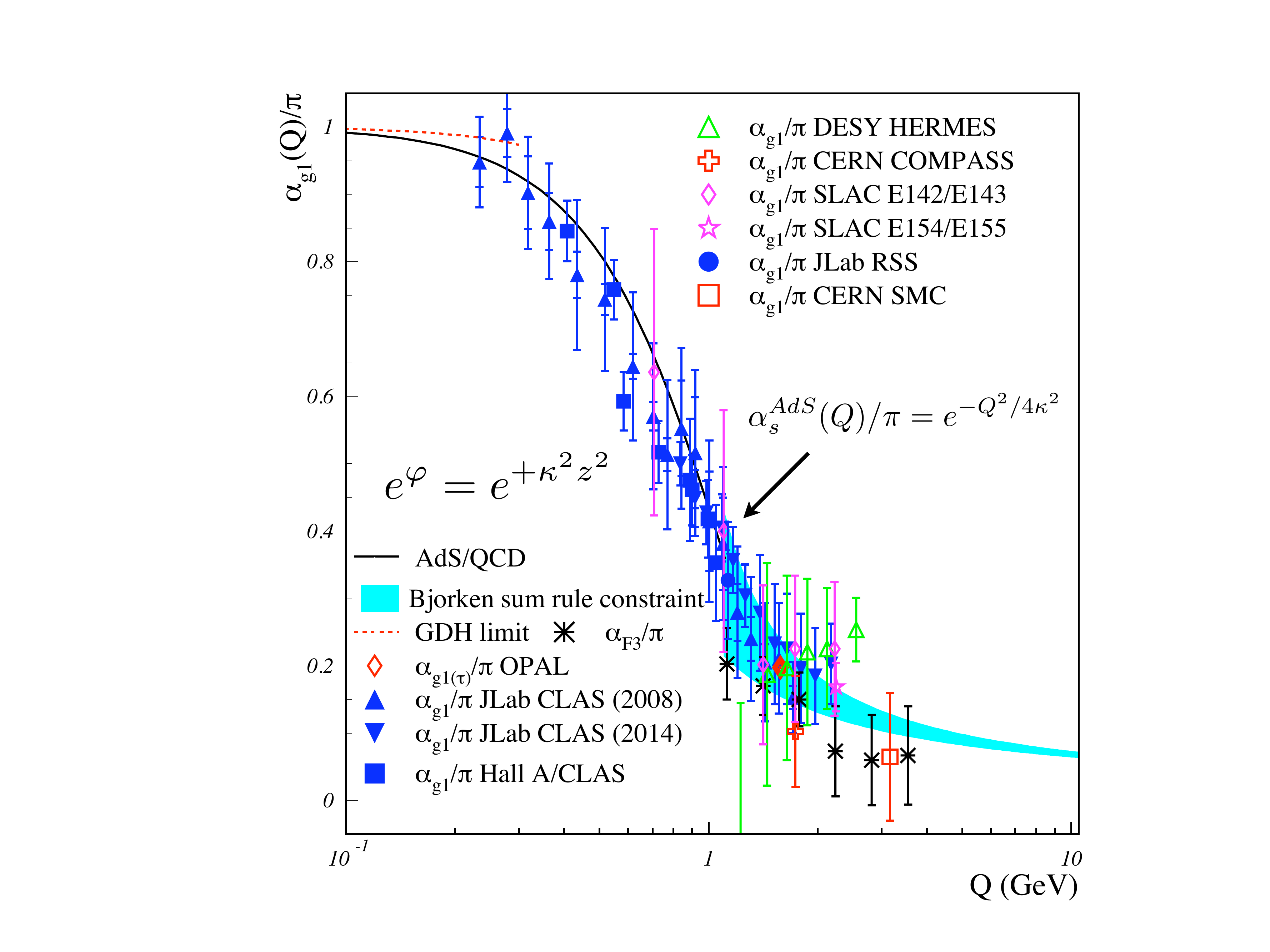}
\includegraphics[height=12cm,width=16cm]{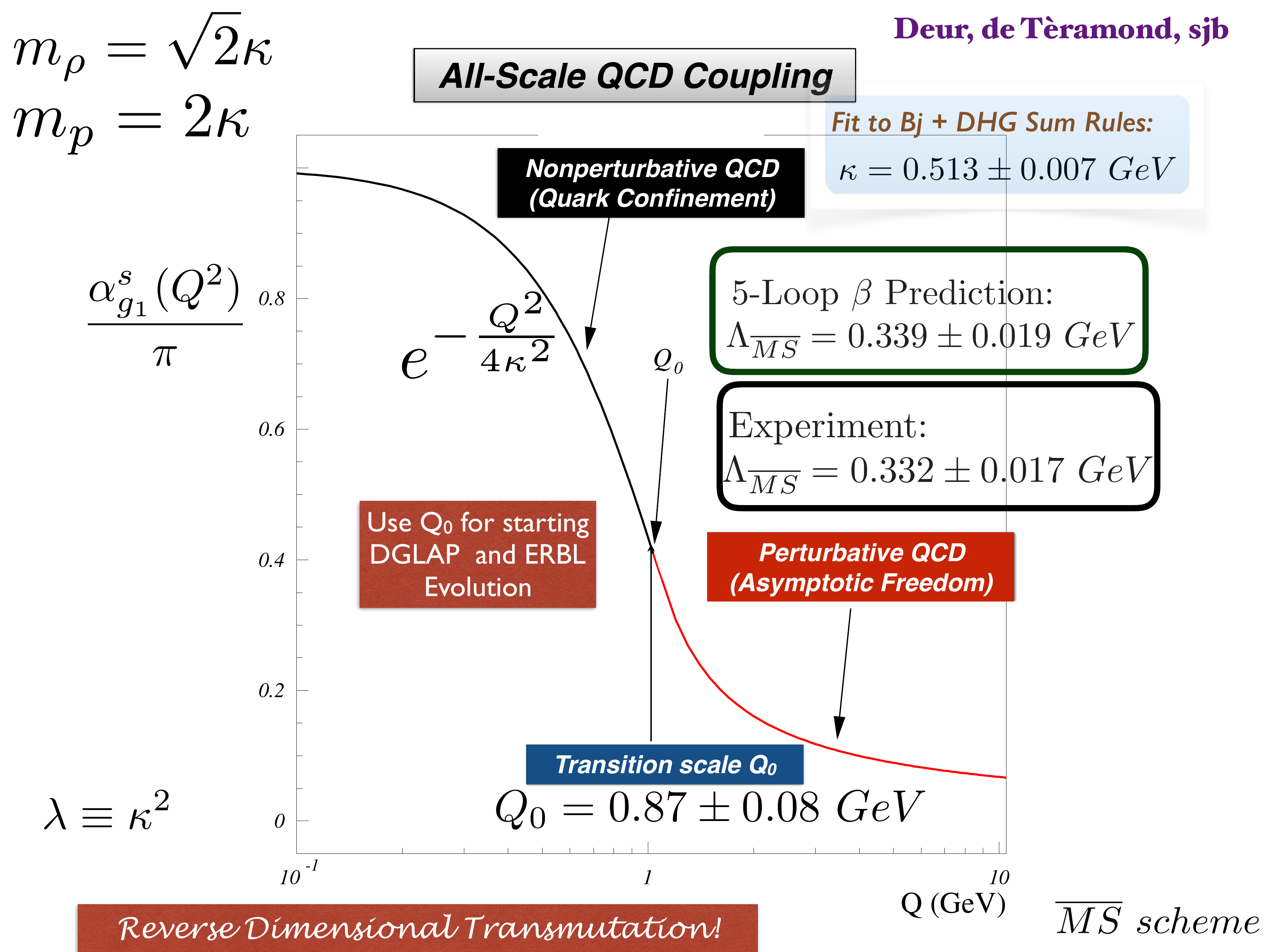}
\end{center}
\caption{
(A) Comparison of the predicted nonpertubative coupling, based on  the dilaton $\exp{(+\kappa^2 z^2)}$ modification of the AdS$_5$ metric, with measurements of the effective charge $\alpha^s_{g_1}(Q^2)$, 
as defined from the Bjorken sum rule.
(B)  Prediction from LF Holography and pQCD for the QCD running coupling $\alpha^s_{g_1}(Q^2)$ at all scales.   The magnitude and derivative of the perturbative and nonperturbative coupling are matched at the scale $Q_0$.  This matching connects the perturbative scale 
$\Lambda_{\overline{MS}}$ to the nonpertubative scale $\kappa$ which underlies the hadron mass scale. 
See Ref.~\cite{Brodsky:2014jia}. 
}
\label{DeurCoupling}
\end{figure}

The dilaton  $e^{+\kappa^2 z^2}$ soft-wall modification of the AdS$_5$ metric, together with LF holography, predicts the functional behavior of the running coupling
in the small $Q^2$ domain~\cite{Brodsky:2010ur}: 
${\alpha^s_{g_1}(Q^2) = 
\pi   e^{- Q^2 /4 \kappa^2 }}. $ 
Measurements of  $\alpha^s_{g_1}(Q^2)$ are remarkably consistent~\cite{Deur:2005cf}  with this predicted Gaussian form; the best fit gives $\kappa= 0.513 \pm 0.007~GeV$.   
See Fig.~\ref{DeurCoupling}.
Deur, de T\'eramond, and I~\cite{Brodsky:2010ur,Deur:2014qfa,Brodsky:2014jia} have also shown how the parameter $\kappa$,  which   determines the mass scale of  hadrons and Regge slopes  in the zero quark mass limit, can be connected to the  mass scale $\Lambda_s$  controlling the evolution of the perturbative QCD coupling.  The high momentum transfer dependence  of the coupling $\alpha_{g1}(Q^2)$ is  predicted  by  pQCD.  The 
matching of the high and low momentum transfer regimes  of $\alpha_{g1}(Q^2)$ -- both its value and its slope -- then determines a scale $Q_0 =0.87 \pm 0.08$ GeV which sets the interface between perturbative and nonperturbative hadron dynamics.  This connection can be done for any choice of renormalization scheme, such as the $\overline{MS}$ scheme,
as seen in  Fig.~\ref{DeurCoupling}.  
The result of this perturbative/nonperturbative matching is an effective QCD coupling  defined at all momenta.   
The predicted value of $\Lambda_{\overline{MS}} = 0.339 \pm 0.019~GeV$ from this analysis agrees well the measured value~\cite{Agashe:2014kda}  
$\Lambda_{\overline{MS}} = 0.332 \pm 0.017~GeV.$
These results, combined with the AdS/QCD superconformal predictions for hadron spectroscopy, allow one to compute hadron masses in terms of $\Lambda_{\overline{MS}}$:
$m_p =  \sqrt 2 \kappa = 3.21~ \Lambda_{\overline{MS}},~ m_\rho = \kappa = 2.2 ~ \Lambda_{\overline{ MS} }, $ and $m_p = \sqrt 2 m_\rho, $ meeting a challenge proposed by Zee~\cite{Zee:2003mt}.
The value of $Q_0$ can be used to set the factorization scale for DGLAP evolution of hadronic structure functions and the ERBL evolution of distribution amplitudes.
Deur, de T\'eramond, and I have also computed the dependence of $Q_0$ on the choice of the  effective charge used to define the running coupling and the renormalization scheme used to compute its behavior in the perturbative regime.   
The use of  the scale $Q_0$  to  resolve  the factorization scale uncertainty in structure functions and fragmentation functions,  in combination with the scheme-independent {\it principle of maximum conformality} (PMC )~\cite{Mojaza:2012mf} for  setting   renormalization scales,  can 
greatly improve the precision of pQCD predictions for collider phenomenology.

\section{Summary} 

QCD is not supersymmetrical in the traditional sense -- the QCD Lagrangian is based on quark and gluonic fields, not squarks nor gluinos.  However its hadronic eigensolutions conform to a supersymmetric representation of superconformal algebra, reflecting the underlying conformal symmetry of chiral QCD and its Pauli matrix representation.   One observes remarkable supersymmetric relations between mesons,  baryons, and tetraquarks  of the same parity as members of the same 4-plet representation of superconformal algebra.   This not only implies identical masses for the bosonic and fermionic hadron eigenvalues, but also supersymmetric relations between their eigenfunctions-- their light-front wavefunctions.  The baryonic eigensolutions correspond to bound states of $3_C$ quarks to a $\bar 3_C$ spin-0 or spin-1 diquark cluster;  the tetraquarks in the 4-plet are bound states of diquarks and  anti-diquarks.  One  predicts universal Regge-slopes in $n$ and $L$ for mesons:   $M^2(n,L) = 4\kappa^2(n+L)$ for mesons and $M^2(n,L) = 4\kappa^2(n+L+1)$ for baryons, consistent with observed hadronic spectroscopy. The pion  eigenstate with $(n=L=S=0)$  thus has zero mass in the chiral $m_q \to 0$ limit.
The supersymmetry of the 4-plet representation is also exhibited dynamically in terms of common features of the light-front wavefunctions of mesons, baryons, and tetraquarks.  Empirically viable predictions for spacelike and timelike hadronic  form factors, structure functions, distribution amplitudes, and transverse momentum distributions have been obtained~\cite{Sufian:2016hwn}.  One can also observe features of superconformal symmetry in the spectroscopy of heavy-light mesons and baryons.

The combined approach of light-front holography and superconformal algebra also provides insight into the origin of the QCD mass scale and color confinement.  A key observation is the remarkable dAFF principle which shows how a mass scale can appear in the Hamiltonian and the equations of motion while retaining the conformal symmetry of the action.  When one applies the dAFF procedure to chiral QCD, a mass scale $\kappa$ appears which determines universal Regge slopes, hadron masses in the absence of the Higgs coupling, and the mass parameter underlying the Gaussian functional form of the nonperturbative QCD running coupling:  $\alpha_s(Q^2) \propto \exp{-(Q^2/4 \kappa^2)}$. This prediction is in agreement with the effective charge  determined from measurements of the Bjorken sum rule.

The potential which underlies color confinement in the effective LF Hamiltonian for the $q \bar q$ Fock state of  mesons  is simply  $U(\zeta^2) = \kappa^4 \zeta^2$, a harmonic oscillator potential in the frame-invariant light-front radial variable $\zeta^2 = b^2_\perp x(1-x)$.  This confinement potential also underlies the spectroscopy and structure of baryons and  
tetraquarks. The same potential can also be derived from the anti--deSitter space representation of the conformal group if the AdS$_5$ is action is modified in the fifth dimension $z$ by the dilaton $e^{+ \kappa^2 z^2}$. This correspondence is based on light-front holography,  the duality between dynamics in physical space-time at fixed LF time and five-dimensional AdS space.  The predicted light-front wavefunctions can also be used to model ``hadronization at the amplitude level".  One can also compute the form of ``direct" subprocesses such as $ u u \to p \bar d$  where a hadron is formed  at high transverse momentum inside the subprocess itself. These processes, together with color transparency,  can explain the ``baryon anomaly" observed in nuclear colisions at RHIC~\cite{Brodsky:2008qp} and the higher-twist scaling of inclusive cross sections such as 
${d\sigma\over d^3p/E}( p p \to H X)$ at fixed $x_T = 2 {p_T\over \sqrt s}$\cite{Arleo:2009ch}.

The mass scale $\kappa$ underlying confinement and hadron masses  can be connected to the parameter   $\Lambda_{\overline {MS}}$ in the QCD running coupling by matching the nonperturbative prediction to the perturbative QCD  regime. The result is an effective coupling  defined at all momenta.   This analysis also determines  $\Lambda_{\overline {MS}}$ in terms of the proton or the $\rho$ mass.    This connection could be implemented in any pQCD renormalization scheme.  
The matching of the high and low momentum transfer regimes also determines a scale $Q_0$ which  sets the interface between perturbative and nonperturbative hadron dynamics.  
The use of $Q_0$ to  resolve  the factorization scale uncertainty for structure functions and distribution amplitudes,  in combination with the  scheme-indepedent {\it Principle of Maximal Conformality (PMC) } for  setting the  renormalization scales~\cite{Mojaza:2012mf},  can 
greatly improve the precision of perturbative QCD predictions for collider phenomenology.  
The absence of vacuum excitations of the causal, frame-independent front form vacuum has important consequences  for the cosmological constant~\cite{Brodsky:2012ku}.

It should be emphasized that the parameter $\kappa$ is not determined in absolute units such as MeV; however,  the ratios of mass parameters such as $m_p/m_\rho = \sqrt 2$ are predicted.   The mass scale $\kappa$ underlying confinement and hadron masses  can be connected to the parameter   $\Lambda_{\overline {MS}}$ in the QCD running coupling by matching the nonperturbative dynamics to the perturbative QCD  regime. This analysis also  gives  a connection between nonperturbative QCD  and PQCD at a scale $Q_0$ and  a prediction for $\Lambda_{\overline {MS}}$ from the proton or $\rho$ mass.

\section*{Acknowledgments}

Presented at NSTAR 2017, 
The 11th International Workshop on the Physics of Excited Nucleons 
August 20 Ñ 23, 2017,
at the University of South Carolina, Columbia, S. C. 
I thank Ralf Gothe for organizing this outstanding meeting.
The physics results presented here are based on collaborations  and discussions  with  Marina Nielsen,
Kelly Chiu, Alexandre Deur, Guy de T\'eramond, Hans G\"unter Dosch, F. G. Cao, G. P. Lepage, 
Rich Lebed, Cedric Lorc\'e, Valery Lyubovitskij, Peter Lowdon,  Raza Sabbir Sufian, S. Glazek, A.  P. Trawinski, 
Matin Mojaza, Ivan Schmidt, Peter Tandy, Francois Arleo, Dae Sung Hwang, Anne Sickles, and Xing-Gang Wu.
This research was supported by the Department of Energy,  contract DE--AC02--76SF00515.  
SLAC-PUB-17201.


\begin{thebibliography}{30}


\bibitem{Klempt:2012fy} 
  E.~Klempt and B.~C.~Metsch,
  Eur.\ Phys.\ J.\ A {\bf 48}, 127 (2012).
  doi:10.1140/epja/i2012-12127-1
  
  
\bibitem{deTeramond:2014asa} 
  G.~F.~de T\'eramond, H.~G.~Dosch and S.~J.~Brodsky,
  Phys.\ Rev.\ D {\bf 91}, no. 4, 045040 (2015)
  [arXiv:1411.5243 [hep-ph]].



\bibitem{Dosch:2015nwa} 
  H.~G.~Dosch, G.~F.~de T\'eramond and S.~J.~Brodsky,
  Phys.\ Rev.\ D {\bf 91}, no. 8, 085016 (2015)
  [arXiv:1501.00959 [hep-th]].



\bibitem{Dosch:2015bca} 
  H.~G.~Dosch, G.~F.~de T\'eramond and S.~J.~Brodsky,
  Phys.\ Rev.\ D {\bf 92}, no. 7, 074010 (2015)
  [arXiv:1504.05112 [hep-ph]].
  
  
  
\bibitem{Haag:1974qh} 
  R.~Haag, J.~T.~Lopuszanski and M.~Sohnius,
  Nucl.\ Phys.\ B {\bf 88}, 257 (1975).
  
  
\bibitem{Brodsky:1980zm} 
  S.~J.~Brodsky and S.~D.~Drell,
  Phys.\ Rev.\ D {\bf 22}, 2236 (1980).
  doi:10.1103/PhysRevD.22.2236
  
  
    
\bibitem{Brodsky:2002cx} 
  S.~J.~Brodsky, D.~S.~Hwang and I.~Schmidt,
  Phys.\ Lett.\ B {\bf 530}, 99 (2002)
  doi:10.1016/S0370-2693(02)01320-5
  [hep-ph/0201296].

  



\bibitem{Nielsen}
 M. Nielsen and  S.~J.~Brodsky (in preparation).


\bibitem{Trawinski:2014msa} 
  A.~P.~Trawinski, S.~D.~Glazek, S.~J.~Brodsky, G.~F.~de T\'eramond and H.~G.~Dosch,
  Phys.\ Rev.\ D {\bf 90}, no. 7, 074017 (2014)
  doi:10.1103/PhysRevD.90.074017
  [arXiv:1403.5651 [hep-ph]].



\bibitem{Sufian:2016hwn} 
  R.~S.~Sufian, G.~F.~de T\'eramond, S.~J.~Brodsky, A.~Deur and H.~G.~Dosch,
  Phys.\ Rev.\ D {\bf 95}, no. 1, 014011 (2017)
  doi:10.1103/PhysRevD.95.014011
  [arXiv:1609.06688 [hep-ph]].
  
\bibitem{deTeramond:2016htp} 
  G.~F.~de T\'eramond, S.~J.~Brodsky, A.~Deur, H.~G.~Dosch and R.~S.~Sufian,
  EPJ Web Conf.\  {\bf 137}, 03023 (2017)
  doi:10.1051/epjconf/201713703023
  [arXiv:1611.03763 [hep-ph]].




	 
\bibitem{deAlfaro:1976je}
 V.~de Alfaro, S.~Fubini and G.~Furlan,
  Nuovo Cim.\ A {\bf 34}, 569 (1976).
  
  
 

\bibitem{Brodsky:2013ar} 
  S.~J.~Brodsky, G.~F.~de T\'eramond and H.~G.~Dosch,
  Phys.\ Lett.\ B {\bf 729}, 3 (2014)
  [arXiv:1302.4105 [hep-th]].
  
\bibitem{Brodsky:2016vig} 
  S.~J.~Brodsky,
  Few Body Syst.\  {\bf 58}, no. 3, 133 (2017)
  doi:10.1007/s00601-017-1292-4
  [arXiv:1611.07194 [hep-ph]].
  
\bibitem{Brodsky:2016nsn} 
  S.~J.~Brodsky,
ÊÊFew Body Syst.\  {\bf 57}, no. 8, 703 (2016)
ÊÊdoi:10.1007/s00601-016-1070-8
ÊÊ[arXiv:1601.06328 [hep-ph]].
ÊÊ
  
  
\bibitem{Brodsky:2017tyf} 
  S.~J.~Brodsky,
ÊÊRuss.\ Phys.\ J.\  {\bf 60}, no. 3, 399 (2017).
ÊÊdoi:10.1007/s11182-017-1089-4
ÊÊ
  




\bibitem{Dirac:1949cp} 
  P.~A.~M.~Dirac,
  Rev.\ Mod.\ Phys.\  {\bf 21}, 392 (1949).



\bibitem{Brodsky:1997de} 
  S.~J.~Brodsky, H.~C.~Pauli and S.~S.~Pinsky,
  Phys.\ Rept.\  {\bf 301}, 299 (1998)
  [hep-ph/9705477].
  
  
\bibitem{Reinhardt:2016fjl} 
  H.~Reinhardt,
  Phys.\ Rev.\ D {\bf 95}, no. 4, 045015 (2017)
  doi:10.1103/PhysRevD.95.045015
  [arXiv:1612.07740 [hep-th]].
  
  
  



\bibitem{Terrell:1959zz} 
  J.~Terrell,
  Phys.\ Rev.\  {\bf 116}, 1041 (1959).



\bibitem{Penrose:1959vz} 
  R.~Penrose,
  Proc.\ Cambridge Phil.\ Soc.\  {\bf 55}, 137 (1959).







\bibitem{Fubini:1984hf} 
  S.~Fubini and E.~Rabinovici,
  Nucl.\ Phys.\ B {\bf 245}, 17 (1984).



  
\bibitem{Munger:1993kq} 
  C.~T.~Munger, S.~J.~Brodsky and I.~Schmidt,
  Phys.\ Rev.\ D {\bf 49}, 3228 (1994).
  doi:10.1103/PhysRevD.49.3228



\bibitem{tHooft:2004doe} 
  G.~'t Hooft,
  hep-th/0408148.



\bibitem{Liu:2015jna} 
  T.~Liu and B.~Q.~Ma,
  Phys.\ Rev.\ D {\bf 92}, no. 9, 096003 (2015)
  [arXiv:1510.07783 [hep-ph]].



\bibitem{Dosch:2016zdv} 
  H.~G.~Dosch, G.~F.~de T\'eramond and S.~J.~Brodsky,
  Phys.\ Rev.\ D {\bf 95}, no. 3, 034016 (2017)
  [arXiv:1612.02370 [hep-ph]].



\bibitem{Brodsky:2000xy} 
  S.~J.~Brodsky, M.~Diehl and D.~S.~Hwang,
  Nucl.\ Phys.\ B {\bf 596}, 99 (2001)
  [hep-ph/0009254].



\bibitem{Brodsky:2000ii} 
  S.~J.~Brodsky, D.~S.~Hwang, B.~Q.~Ma and I.~Schmidt,
  Nucl.\ Phys.\ B {\bf 593}, 311 (2001)
  [hep-th/0003082].



\bibitem{Kobzarev:1962wt} 
  I.~Y.~Kobzarev and L.~B.~Okun,
  Zh.\ Eksp.\ Teor.\ Fiz.\  {\bf 43}, 1904 (1962)
  [Sov.\ Phys.\ JETP {\bf 16}, 1343 (1963)].



\bibitem{Teryaev:1999su} 
  O.~V.~Teryaev,
  hep-ph/9904376.



\bibitem{Brodsky:2011xx} 
  S.~J.~Brodsky, F.~G.~Cao and G.~F.~de T\'eramond,
  Phys.\ Rev.\ D {\bf 84}, 075012 (2011)
  [arXiv:1105.3999 [hep-ph]].



\bibitem{Forshaw:2012im} 
  J.~R.~Forshaw and R.~Sandapen,
  Phys.\ Rev.\ Lett.\  {\bf 109}, 081601 (2012)
  [arXiv:1203.6088 [hep-ph]].



\bibitem{deTeramond:2013it} 
  G.~F.~de T\'eramond, H.~G.~Dosch and S.~J.~Brodsky,
  Phys.\ Rev.\ D {\bf 87}, no. 7, 075005 (2013)
  [arXiv:1301.1651 [hep-ph]].



\bibitem{Brodsky:2014yha} 
  S.~J.~Brodsky, G.~F.~de T\'eramond, H.~G.~Dosch and J.~Erlich,
  Phys.\ Rept.\  {\bf 584}, 1 (2015)
  [arXiv:1407.8131 [hep-ph]].



\bibitem{Gutsche:2015xva} 
  T.~Gutsche, V.~E.~Lyubovitskij, I.~Schmidt and A.~Vega,
  Phys.\ Rev.\ D {\bf 91}, no. 11, 114001 (2015)
  [arXiv:1501.02738 [hep-ph]].



\bibitem{Gutsche:2016lrz} 
  T.~Gutsche, V.~E.~Lyubovitskij and I.~Schmidt,
  Phys.\ Rev.\ D {\bf 94}, no. 11, 116006 (2016)
  [arXiv:1607.04124 [hep-ph]].



\bibitem{deTeramond:2008ht} 
  G.~F.~de T\'eramond and S.~J.~Brodsky,
  Phys.\ Rev.\ Lett.\  {\bf 102}, 081601 (2009)
  [arXiv:0809.4899 [hep-ph]].



\bibitem{Smirnov:2009fh} 
  A.~V.~Smirnov, V.~A.~Smirnov and M.~Steinhauser,
  Phys.\ Rev.\ Lett.\  {\bf 104}, 112002 (2010)
  [arXiv:0911.4742 [hep-ph]].



\bibitem{Ashery:2000yj} 
  D.~Ashery,
  Nucl.\ Phys.\ Proc.\ Suppl.\  {\bf 90}, 67 (2000)
  [Nucl.\ Phys.\ Proc.\ Suppl.\  {\bf 108}, 321 (2002)]
  [hep-ex/0008036].
  
\bibitem{Reinhardt:2012xs} 
  H.~Reinhardt and H.~Weigel,
  Phys.\ Rev.\ D {\bf 85}, 074029 (2012)
  doi:10.1103/PhysRevD.85.074029
  [arXiv:1201.3262 [hep-ph]].
  
  
\bibitem{Srivastava:2002mw} 
  P.~P.~Srivastava and S.~J.~Brodsky,
  Phys.\ Rev.\ D {\bf 66}, 045019 (2002)
  [hep-ph/0202141].



\bibitem{Gribov:1972ri} 
  V.~N.~Gribov and L.~N.~Lipatov,
  Sov.\ J.\ Nucl.\ Phys.\  {\bf 15}, 438 (1972)
  [Yad.\ Fiz.\  {\bf 15}, 781 (1972)].



\bibitem{Altarelli:1977zs} 
  G.~Altarelli and G.~Parisi,
  Nucl.\ Phys.\ B {\bf 126}, 298 (1977).



\bibitem{Dokshitzer:1977sg} 
  Y.~L.~Dokshitzer,
  Sov.\ Phys.\ JETP {\bf 46}, 641 (1977)
  [Zh.\ Eksp.\ Teor.\ Fiz.\  {\bf 73}, 1216 (1977)].



\bibitem{Lepage:1979zb} 
  G.~P.~Lepage and S.~J.~Brodsky,
  Phys.\ Lett.\  {\bf 87B}, 359 (1979).



\bibitem{Lepage:1980fj} 
  G.~P.~Lepage and S.~J.~Brodsky,
  Phys.\ Rev.\ D {\bf 22}, 2157 (1980).



\bibitem{Efremov:1979qk} 
  A.~V.~Efremov and A.~V.~Radyushkin,
  Phys.\ Lett.\  {\bf 94B}, 245 (1980).



\bibitem{Efremov:1978rn} 
  A.~V.~Efremov and A.~V.~Radyushkin,
  Theor.\ Math.\ Phys.\  {\bf 42}, 97 (1980)
  [Teor.\ Mat.\ Fiz.\  {\bf 42}, 147 (1980)].



\bibitem{Brodsky:2015uwa} 
  S.~J.~Brodsky and S.~Gardner,
  Phys.\ Rev.\ Lett.\  {\bf 116}, no. 1, 019101 (2016)
  [arXiv:1504.00969 [hep-ph]].



\bibitem{Pauli:1985pv} 
  H.~C.~Pauli and S.~J.~Brodsky,
  Phys.\ Rev.\ D {\bf 32}, 1993 (1985).



\bibitem{Hornbostel:1988fb} 
  K.~Hornbostel, S.~J.~Brodsky and H.~C.~Pauli,
  Phys.\ Rev.\ D {\bf 41}, 3814 (1990).



\bibitem{Vary:2014tqa} 
  J.~P.~Vary, X.~Zhao, A.~Ilderton, H.~Honkanen, P.~Maris and S.~J.~Brodsky,
  Nucl.\ Phys.\ Proc.\ Suppl.\  {\bf 251-252}, 10 (2014)
  [arXiv:1406.1838 [nucl-th]].



\bibitem{Brodsky:2015aia} 
  S.~J.~Brodsky {\it et al.},
  arXiv:1502.05728 [hep-ph].



\bibitem{Brodsky:2009gx} 
  S.~J.~Brodsky and R.~F.~Lebed,
  Phys.\ Rev.\ Lett.\  {\bf 102}, 213401 (2009)
  [arXiv:0904.2225 [hep-ph]].



\bibitem{Banburski:2012tk} 
  A.~Banburski and P.~Schuster,
  Phys.\ Rev.\ D {\bf 86}, 093007 (2012)
  [arXiv:1206.3961 [hep-ph]].



\bibitem{Vary:2013kma} 
  J.~P.~Vary, X.~Zhao, A.~Ilderton, H.~Honkanen, P.~Maris and S.~J.~Brodsky,
  Acta Phys.\ Polon.\ Supp.\  {\bf 6}, 257 (2013).



\bibitem{Ashery:2005wa} 
  D.~Ashery,
  Nucl.\ Phys.\ Proc.\ Suppl.\  {\bf 161}, 8 (2006)
  [hep-ex/0511052].



\bibitem{Brodsky:1983vf} 
  S.~J.~Brodsky, C.~R.~Ji and G.~P.~Lepage,
  Phys.\ Rev.\ Lett.\  {\bf 51}, 83 (1983).



\bibitem{Cruz-Santiago:2015dla} 
  C.~Cruz-Santiago, P.~Kotko and A.~M.~Sta?to,
  Prog.\ Part.\ Nucl.\ Phys.\  {\bf 85}, 82 (2015).





\bibitem{Brodsky:1973kb} 
  S.~J.~Brodsky, R.~Roskies and R.~Suaya,
  Phys.\ Rev.\ D {\bf 8}, 4574 (1973).



\bibitem{Brodsky:2009dr} 
  S.~J.~Brodsky and G.~F.~de T\'eramond,
  arXiv:0901.0770 [hep-ph].



\bibitem{Chiu:2017ycx} 
  K.~Y.~J.~Chiu and S.~J.~Brodsky,
  Phys.\ Rev.\ D {\bf 95}, no. 6, 065035 (2017)
  [arXiv:1702.01127 [hep-th]].



\bibitem{Jaffe:1987sx} 
  R.~L.~Jaffe,
  Phys.\ Lett.\ B {\bf 193}, 101 (1987).



\bibitem{Brodsky:2012ku} 
  S.~J.~Brodsky, C.~D.~Roberts, R.~Shrock and P.~C.~Tandy,
  Phys.\ Rev.\ C {\bf 85}, 065202 (2012)
  [arXiv:1202.2376 [nucl-th]].



\bibitem{Zee:2008zz} 
  A.~Zee,
  Mod.\ Phys.\ Lett.\ A {\bf 23}, 1336 (2008).



\bibitem{Casher:1974xd} 
  A.~Casher and L.~Susskind,
  Phys.\ Rev.\ D {\bf 9}, 436 (1974).



\bibitem{Brodsky:2009zd} 
  S.~J.~Brodsky and R.~Shrock,
  Proc.\ Nat.\ Acad.\ Sci.\  {\bf 108}, 45 (2011)
  [arXiv:0905.1151 [hep-th]].



\bibitem{Brodsky:2010xf} 
  S.~J.~Brodsky, C.~D.~Roberts, R.~Shrock and P.~C.~Tandy,
  Phys.\ Rev.\ C {\bf 82}, 022201 (2010)
  [arXiv:1005.4610 [nucl-th]].





\bibitem{Verlinde:2016toy} 
  E.~P.~Verlinde,
  arXiv:1611.02269 [hep-th].



\bibitem{Grunberg:1980ja} 
  G.~Grunberg,
  Phys.\ Lett.\  {\bf 95B}, 70 (1980)
  Erratum: [Phys.\ Lett.\  {\bf 110B}, 501 (1982)].



\bibitem{Brodsky:1994eh} 
  S.~J.~Brodsky and H.~J.~Lu,
  Phys.\ Rev.\ D {\bf 51}, 3652 (1995)
  [hep-ph/9405218].



\bibitem{Brodsky:2010ur} 
  S.~J.~Brodsky, G.~F.~de T\'eramond and A.~Deur,
  Phys.\ Rev.\ D {\bf 81}, 096010 (2010)
  [arXiv:1002.3948 [hep-ph]].



\bibitem{Deur:2005cf} 
  A.~Deur, V.~Burkert, J.~P.~Chen and W.~Korsch,
  Phys.\ Lett.\ B {\bf 650}, 244 (2007)
  [hep-ph/0509113].



\bibitem{Deur:2014qfa} 
  A.~Deur, S.~J.~Brodsky and G.~F.~de T\'eramond,
  Phys.\ Lett.\ B {\bf 750}, 528 (2015)
  [arXiv:1409.5488 [hep-ph]].



\bibitem{Brodsky:2014jia} 
  S.~J.~Brodsky, G.~F.~de T\'eramond, A.~Deur and H.~G.~Dosch,
  Few Body Syst.\  {\bf 56}, no. 6-9, 621 (2015)
  [arXiv:1410.0425 [hep-ph]].



\bibitem{Agashe:2014kda} 
  K.~A.~Olive {\it et al.} [Particle Data Group],
  Chin.\ Phys.\ C {\bf 38}, 090001 (2014).



\bibitem{Zee:2003mt} 
  A.~Zee,
  Princeton, UK: Princeton Univ. Pr. (2010) 576 p



\bibitem{Mojaza:2012mf} 
  M.~Mojaza, S.~J.~Brodsky and X.~G.~Wu,
  Phys.\ Rev.\ Lett.\  {\bf 110}, 192001 (2013)
  [arXiv:1212.0049 [hep-ph]].







\bibitem{Liuti:2013cna} 
  S.~Liuti, A.~Rajan, A.~Courtoy, G.~R.~Goldstein and J.~O.~Gonzalez Hernandez,
  Int.\ J.\ Mod.\ Phys.\ Conf.\ Ser.\  {\bf 25}, 1460009 (2014)
  [arXiv:1309.7029 [hep-ph]].



\bibitem{Mondal:2015uha} 
  C.~Mondal and D.~Chakrabarti,
  Eur.\ Phys.\ J.\ C {\bf 75}, no. 6, 261 (2015)
  [arXiv:1501.05489 [hep-ph]].



\bibitem{Lorce:2011dv} 
  C.~Lorce, B.~Pasquini and M.~Vanderhaeghen,
  JHEP {\bf 1105}, 041 (2011)
  [arXiv:1102.4704 [hep-ph]].



\bibitem{Brodsky:2008xe} 
  S.~J.~Brodsky,
  AIP Conf.\ Proc.\  {\bf 1105}, 315 (2009)
  [arXiv:0811.0875 [hep-ph]].



\bibitem{Brodsky:2009dv} 
  S.~J.~Brodsky,
  Nucl.\ Phys.\ A {\bf 827}, 327C (2009)
  [arXiv:0901.0781 [hep-ph]].





\bibitem{Brodsky:2002ue} 
  S.~J.~Brodsky, P.~Hoyer, N.~Marchal, S.~Peigne and F.~Sannino,
  Phys.\ Rev.\ D {\bf 65}, 114025 (2002)
  [hep-ph/0104291].



\bibitem{Brodsky:2010vs} 
  S.~J.~Brodsky, B.~Pasquini, B.~W.~Xiao and F.~Yuan,
  Phys.\ Lett.\ B {\bf 687}, 327 (2010)
  [arXiv:1001.1163 [hep-ph]].



\bibitem{Brodsky:2013oya} 
  S.~J.~Brodsky, D.~S.~Hwang, Y.~V.~Kovchegov, I.~Schmidt and M.~D.~Sievert,
  Phys.\ Rev.\ D {\bf 88}, no. 1, 014032 (2013)
  [arXiv:1304.5237 [hep-ph]].



\bibitem{Brodsky:1989qz} 
  S.~J.~Brodsky and H.~J.~Lu,
  Phys.\ Rev.\ Lett.\  {\bf 64}, 1342 (1990).



\bibitem{Brodsky:2004qa} 
  S.~J.~Brodsky, I.~Schmidt and J.~J.~Yang,
  Phys.\ Rev.\ D {\bf 70}, 116003 (2004)
  [hep-ph/0409279].



\bibitem{Schienbein:2007fs} 
  I.~Schienbein, J.~Y.~Yu, C.~Keppel, J.~G.~Morfin, F.~Olness and J.~F.~Owens,
  Phys.\ Rev.\ D {\bf 77}, 054013 (2008)
  [arXiv:0710.4897 [hep-ph]].



  
\bibitem{Brodsky:2008qp} 
  S.~J.~Brodsky and A.~Sickles,
  Phys.\ Lett.\ B {\bf 668}, 111 (2008)
  doi:10.1016/j.physletb.2008.07.108
  [arXiv:0804.4608 [hep-ph]].
  
\bibitem{Arleo:2009ch} 
  F.~Arleo, S.~J.~Brodsky, D.~S.~Hwang and A.~M.~Sickles,
  Phys.\ Rev.\ Lett.\  {\bf 105}, 062002 (2010)
  doi:10.1103/PhysRevLett.105.062002
  [arXiv:0911.4604 [hep-ph]].



\end{thebibliography}
\end{document}